\documentclass{JHEP3}

\usepackage{Pformel,Pmeta}
\usepackage{amsmath,amsfonts,amsthm,times}
\usepackage{pstricks,Ppicture}
\usepackage{multirow}

\title{Conformal Toda theory with a boundary}

\author{Vladimir Fateev$^{1,2}$
and Sylvain Ribault$^1$ 
\\ \!\!\!$^1$\!
 Laboratoire de Physique Th\'eorique et Astroparticules, UMR5207 CNRS-UM2,
 \\
 Universit\'e Montpellier II, Place E. Bataillon,
 34095 Montpellier Cedex 05, France 
 \\
\!\!\!$^2$\! Landau Institute for Theoretical Physics
 \\
 142432 Chernogolovka, Russia
 \\
 {\footnotesize \tt vladimir.fateev@lpta.univ-montp2.fr, sribault@um2.fr }
}

\abstract{We investigate $s\ell_n$ conformal Toda theory with maximally symmetric boundaries. There are two types of maximally symmetric boundary conditions, due to the existence of an order two automorphism of the $W_{n\geq 3}$ algebra. In one of the two cases, we find that there exist D-branes of all possible dimensions $0\leq d\leq n-1$, which correspond to partly degenerate representations of the $W_n$ algebra. We perform classical and conformal bootstrap analyses of such D-branes, and relate these two approaches by using the semi-classical light asymptotic limit. In particular we determine the bulk one-point functions. We observe remarkably severe divergences in the annulus partition functions, and attribute their origin to the existence of infinite multiplicities in the fusion of representations of the $W_{n\geq 3}$ algebra. We also comment on the issue of the existence of a boundary action, using the calculus of constrained functional forms, and derive the generating function of the B\"acklund transformation for $s\ell_3$ Toda classical mechanics, using the minisuperspace limit of the bulk one-point function.
}

\makeatletter\let\default@color\current@color\makeatother 

\begin{document}

\zeq\section{Introduction}

There are good reasons for studying $s\ell_n$ conformal Toda theories, as in principle these non-rational two-dimensional conformal field theories have all the usual applications of two-dimensional CFTs, applications to quantum gravity, string theory and critical phenomena. (See \cite{fl07c} for more details and references.)
In particular, the simplest and well-studied case of Liouville theory ($n=2$) is an essential tool in the study of non-critical string theories and two-dimensional quantum gravity. And the study of Liouville theory with a boundary plays an important role in the understanding of non-critical open strings and of the corresponding D-branes, which 
account for the non-perturbative effects in non-critical string theory. The other cases ($n\geq 3$) are directly related to the so-called $W$-strings and $W$-gravity theories, whose names come from the $W_n$ symmetry algebra of $s\ell_n$ conformal Toda theory. And the non-perturbative effects in $W$ string theory are expected to be due to D-branes, which can be technically described using conformal Toda theory with a boundary. In addition to such applications, another motivation 
for investigating the higher Toda theories is their beautiful, intricate and challenging nature, which suggests that their study can reveal qualitatively new structures and phenomena in two-dimensional conformal field theory.

The $W_n$ algebra, which is an extension of the Virasoro algebra, was discovered \cite{zam85} \cite{fl88} soon after the seminal work of Belavin, Polyakov and Zamolodchikov on two-dimensional CFTs \cite{bpz84}, and rational CFTs with $W_n$ symmetries were then constructed \cite{fz86b} \cite{fl88}. The study of $s\ell_n$ conformal Toda theories, which are non-rational CFTs with $W_n$ symmetries, is much more recent \cite{fl07c} \cite{fl08}. The case of Liouville theory had to be studied first, and $s\ell_{n\geq 3}$ Toda theory is considerably more complicated than Liouville theory. The reasons for these extra complications can be found in the properties of the $W_n$ algebras, as we will demonstrate. 

Our most powerful tool in the study of non-rational CFTs is the conformal bootstrap method, which purposes to determine all correlation functions once the spectrum of the theory is given, and the $W_n$ symmetry of the theory is assumed. So far this has been achieved only in the case of Liouville theory; however this is in principle doable also in $s\ell_{n\geq 3}$ Toda theories. The conformal bootstrap equations for say the three-point functions are vastly overdetermined, the problem is to find closed subsystems of manageable numbers of equations. In the present article we will achieve this in the case of the one-point function in the presence of a boundary. Introducing a boundary in the two-dimensional space on which our field theory lives of course makes the theory more complicated, but the advantage is that simple correlation functions like the one point function, which has to vanish in the absence of a boundary, now become interesting observables. 

In the case of CFTs with boundaries, the fundamental relations between the properties of the symmetry algebra and the physical observables of the theory were discovered by Cardy \cite{car89},
and we will refer to them as ``Cardy's ideas''. First of all, maximally symmetric D-branes are related to the representations of the symmetry algebra which appear in the bulk spectrum. Then, the spectrum of open strings with their ends on two D-branes is given by the fusion product of the two corresponding representations. We will find that these properties mostly hold in $s\ell_n$ conformal Toda theory, in the cases where we can determine the relevant objects. There will be restrictions, some of which were already observed in the case of Liouville theory: there exist not only continuous D-branes associated to the continuous representations which do appear in the bulk spectrum, but also discrete D-branes associated to degenerate representations which do not. 

So we will begin with a study of $W_n$ algebras and their representations (Section \ref{secw}), where we will emphasize the features which will play an important role in $s\ell_n$ Toda theory: the properties of the characters, the existence of an order $2$ automorphism, the existence of infinite fusion multiplicities, the existence of a hierarchy of partly degenerate representations. We hope that this review will be enough for understanding the rest of the article, but we also recommend the reviews \cite{bs92}\cite{fl90} on $W_n$ algebras and \cite{fms97} on conformal field theory. Then, we will solve the classical Toda equations on the disc (Section \ref{sted}). The resulting picture of the moduli spaces of D-branes will turn out to be qualitatively correct, as will be confirmed by the conformal bootstrap analysis (Section \ref{secbo}). There, the analysis of the differential equations obeyed by certain two-point functions will result in explicit expressions for the one-point functions, which characterize how D-branes couple to bulk operators. The calculation of annulus partition functions will also provide some information on the boundary sector. The relation between the classical and bootstrap analyses will be made precise thanks to the light asymptotic limit (Section \ref{secla}), which will also allow us to predict some correlation functions which are at present out of reach of the bootstrap analysis. The conclusion (Section \ref{secco}) will summarize the main results and remaining puzzles. Then come two Appendices, which are devoted to interesting but peripheral topics: Appendix \ref{secmi} to the minisuperspace limit, which will turn out to lead to the determination of the generating function of the B\"acklund transformation which relates $s\ell_3$ Toda classical mechanics to a free system, and Appendix \ref{secac} to the existence of boundary actions, which we will be able to predict or rule out based on the properties of the boundary conditions.

\zeq\section{$W_n$ algebras and their representations \label{secw}}

The symmetry algebra of the $s\ell_n$ conformal Toda theory is the so-called $W_n$ algebra. The Virasoro algebra coincides with the $W_2$ algebra, and is a subalgebra of the $W_{n>2}$ algebra, so that $s\ell_n$ conformal Toda theory indeed has conformal symmetry. The spectrum of the theory decomposes into representations of the $W_n$ algebra, which we will therefore study. 

The infinite-dimensional $W_n$ algebra is related to the finite-dimensional $s\ell_n$ algebra in a number of ways.
For example,
the Virasoro algebra can be obtained from the affine extension $\asl$ of the $s\ell_2$ algebra by a quantum Hamiltonian reduction of the Drinfeld-Sokolov type.  The $W_n$ algebra can similarly be obtained from $\widehat{s\ell_n}$.
Moreover,
a fully degenerate representation of $W_n$ can be associated to each pair of two highest-weight representations of $s\ell_n$.
This is our motivation for reviewing the representations of $s\ell_n$ (more on this in \cite{fms97}), as an introduction to the study of representations of $W_n$.

\subsection{Representation theory of $s\ell_n$ Lie algebras}

Representations of $s\ell_n$ are parametrized by vectors in an $n-1$-dimensional space spanned by the simple roots $e_1\cdots e_{n-1}$ whose scalar products $K_{i,j}=(e_i,e_j)$ form the Cartan matrix, whose only nonzero entries are $K_{ii}=2,K_{i,i-1}=K_{i,i+1}=-1$. The $\frac12 n(n-1)$ positive roots are the sums of any numbers of consecutive simple roots, in the $s\ell_3$ case they are $\{e>0\}=\{e_1,e_2,e_1+e_2\}$. 
The fundamental weights are the vectors $\omega_i$ such that $(\omega_i,e_j)=\delta_{ij}$. The Weyl vector is
\bea
\rho=\tfrac12 \sum_{e>0} e\ ,
\eea 
and $\rho^2\equiv (\rho,\rho)=\frac{1}{12} (n-1)n(n+1)$. In the $s\ell_3$ case we have 
\bea
\left\{\begin{array}{l} \om_1=\tfrac23 e_1+\tfrac 13 e_2 \\ \om_2=\tfrac13 e_1+\tfrac23 e_2 \end{array}\right.,\
\left\{\begin{array}{l} e_1=2\om_1-\om_2 \\ e_2=2\om_2-\om_1 \end{array}\right.,\
K=\bsm 2 & -1 \\ -1 & 2 \esm,\
\{e>0\}=\{e_1,e_2,\rho\}\ .
\eea
The Weyl group, a finite group, acts on the root space while preserving the scalar product. In the case of $s\ell_2$ it is a $\Z_2$ group whose nontrivial element is the reflection $r(v)=-v$. In the case of $s\ell_3$ the Weyl group has six elements $\{1,r,s,rs,sr,rsr=srs\}$ and can be identified with the group of permutations of the three elements $\{h_i\}\equiv\{\om_1,\om_2-\om_1,-\om_2\}$ with the action
\bea
\begin{array}{|c|c|c|c|c|c|}
 \hline
  1 & r & s & rs & sr & rsr 
 \\
 \hline \hline
  e_1 & \rho & -e_1 & -\rho & e_2 & -e_2
 \\
  e_2 & -e_2 & \rho & e_1 & -\rho & -e_1
 \\
  \rho & e_1 & e_2 & -e_2 & -e_1 & -\rho
 \\
 \hline
 \om_1 &\om_1 &\om_2-\om_1 & -\om_2 & \om_2-\om_1 & -\om_2
 \\
\om_2-\om_1 & -\om_2 & \om_1 & \om_1 & -\om_2 & \om_2-\om_1
 \\
  -\om_2& \om_2-\om_1 & -\om_2 & \om_2-\om_1 & \om_1 & \om_1
 \\
 \hline 
\end{array}
\eea
In the general $s\ell_n$ case, the Weyl group is generated by the $n-1$ reflections $s_i$ such that $s_i(e_j)=e_j-K_{ji}e_i$. The signature of an element of the group is the function $\e$ such that $\e(s_i)=-1$ and $\e(ww')=\e(w)\e(w')$. 

To an integral dominant weight, that is a vector $\Omega=\sum_i \lambda_i \om_i\in \sum_i\N\om_i$, we can associate a finite-dimensional irreducible representation $R_\Omega$ of $s\ell_n$. The vector $\Omega$ is then called its highest weight. A finite number of weights $h\in H_\Omega$ such that $\Omega-h\in \sum_i \N e_i$ are associated to the representation. The weights are the eigenvalues of the generators of the Cartan subalgebra when acting on a basis of the representation, so that the number of weights, taking into account their possible integer multiplicities, is the dimension of the representation. For example, the fundamental representation of $s\ell_n$ has dimension $n$ and weights $H_{\om_1}=\{h_k=\om_1-\sum_{i=1}^k e_i|k=0\cdots n-1\}$. The adjoint representation of $s\ell_3$ has dimension $8$ and weights $H_\rho = \{\pm e_1,\pm e_2,\pm \rho,2\cdot 0\}$ where the weight $0$ appears with multiplicity $2$. Multiplicities higher than one appear only in the cases $s\ell_{n\geq 3}$. 

The character $\chi_\Omega(p)$ of a representation is defined as a function of a vector $p$ by 
\bea
\chi_\Omega(p)\equiv\sum_{h\in H_\Omega} e^{(h,p)} \ . 
\label{chio}
\eea
Given the highest weight $\Omega$ of a representation, the other weights can be found thanks to the Weyl formula
\bea
\chi_\Omega(p) = \frac{\sum_{w\in W} \e(w) e^{(\rho+\Omega,w(p))}}{\sum_{w\in W} \e(w) e^{(\rho,w(p))}}\ ,
\label{cop}
\eea
whose denominator can be rewritten as
\bea
\sum_{w\in W} \e(w) e^{(\rho,w(p))} = \prod_{e>0} (e^{\frac12(e,p)}-e^{-\frac12(e,p)})\ .
\label{swpe}
\eea
Characters behave nicely under tensor products of representations, thanks to the property
\bea
R_\Omega \otimes R_{\Omega'} = \sum_{\Omega''} m^{\Omega''}_{\Omega,\Omega'} R_{\Omega''}\ \Rightarrow\ \chi_\Omega(p)\chi_{\Omega'}(p) = \sum_{\Omega''} m^{\Omega''}_{\Omega,\Omega'} \chi_{\Omega''}(p)\ .
\label{rrr}
\eea
The hyperplanes $\{(e,p)=0\}_{e>0}$ divide the $p$-space into $n!$ Weyl chambers, which are fundamental domains for the action of the Weyl group.

For $n\geq 3$, the algebra $s\ell_n$ has an order two automorphism, called
the Dynkin diagram automorphism, which maps $R_{\Omega}$ to $R_{\Omega^*}$, where the conjugation $\Omega \rar \Omega^*$ is the linear map characterized by $e_i^*=e_{n-i}$. This map is trivial in the case of $s\ell_2$. We will see that this automorphism induces an automorphism of the $W_{n\geq 3}$ algebra.

\subsection{Representation theory of $W_n$ algebras \label{ssrep}}

The algebra $W_n$ is generated by $n-1$ operators $W^{(2)},W^{(3)},\cdots W^{(n)}$, where $W^{(2)}=T$ is the stress-energy tensor. (See the review \cite{bs92}.) Let us explicitly write Zamolodchikov's $W_3$ algebra, where for simplicity we denote $W^{(3)}=W$:
\bea
T(z)T(w)&=&\frac{c/2}{(z-w)^4}+\frac{2T(w)}{(z-w)^2}+\frac{\p T(w)}{z-w}+{\cal O}(1)\ ,
\\
T(z)W(w)&=&\frac{3W(w)}{(z-w)^2} + \frac{ \p W(w)}{(z-w)}+{\cal O}(1)\ ,
\\
W(z)W(w)&=& \frac{c/3}{(z-w)^6} +\frac{2T(w)}{(z-w)^4} + \frac{\p T(w)}{(z-w)^3} + \frac{1}{(z-w)^2}\left[2\beta\Lambda(w)+\frac{3}{10} \p^2T(w)\right]
\nn
\\ & &\hspace{4cm}
 +\frac{1}{z-w}\left[\beta \p \Lambda(w) +\frac{1}{15}\p^3 T(w)\right]+{\cal O}(1)\ ,
\eea
where $\Lambda(w)=(TT)(w)-\frac{3}{10}\p^2 T(w)$ and $\beta=\frac{16}{22+5c}$. The algebra depends on a central charge $c$, which we parametrize in terms of a real number $b$ as $c=(n-1)(1+n(n+1)(b+b^{-1})^2)$. The generators of the algebra can be decomposed into modes $W_n^{(s)}$ as $W^{(s)}(z)=\sum_{n\in\Z} W_n^{(s)}z^{-n-s}$; there is a special notation $L_n$ for the modes of $T(z)=\sum_{n\in \Z} L_n z^{-n-2}$.

A representation of the $W_n$ algebra can be encoded in a vertex operator $V(z)$, and the action of the algebra is encoded in the operator product $W^{(s)}(z) V(w) = \sum_{n\in \Z} \frac{W^{(s)}_n V(w)}{(z-w)^{s-n}}$. 
A standard assumption in conformal field theory is that the spectrum is a sum of highest-weight representations, generated by primary operators such that $W^{(s)}_{n>0} V(w)=0$ and $W^{(s)}_0 V(w)=q^{(s)} V(w)$. The product of a generator $W^{(s)}$ with a primary operator therefore contains a finite number of singular terms. In the $W_3$ case a primary operator $V(w)$ obeys
\bea
T(z) V(w) &=& \frac{\Delta V(w)}{(z-w)^2} +\frac{\p V(w)}{z-w} + {\cal O}(1)\ ,
\label{tv}
\\
W(z) V(w) &=& \frac{q V(w)}{(z-w)^3} +\frac{W_{-1}V(w)}{(z-w)^2} + \frac{W_{-2} V(w)}{z-w} +{\cal O}(1)\ , 
\label{wv}
\eea
where we denote $\Delta=q^{(2)}$ the conformal dimension and $q=q^{(3)}$ the $W$-charge, and we use the identification of $L_{-1}$ with the generator of translations $\p$, which is another standard assumption in conformal field theory. All operators of interest are assumed to be linear combinations of operators of the type $D V(w)=(\prod_{i=1}^N W^{(s_i)}_{-n_i}) V(w)$ where $V(w)$ is primary and $n_i>0, N\geq 0$. A descendent operator of level $L>0$ is a linear combination of such operators with $\sum_i n_i=L$. A descendent which is itself primary is called a null vector. 

A primary operator is in principle characterized by the corresponding $W^{(s)}_0$ eigenvalues $q^{(s)}$, but it is convenient to introduce a redundant parametrization of these eigenvalues and to label operators by an $(n-1)$-dimensional vector $\al$ called the momentum,
\bea
\al=\sum_i \al_i\om_i \quad {\rm so\ that} \quad \al_i=(e_i,\al)\ .
\eea 
The corresponding conformal dimension is supposed to be 
\bea
q^{(2)}_\al = \Delta_\al = \frac12(\al,2Q-\al)\ ,
\label{da}
\eea
where we introduce the vector
\bea
Q=(b+b^{-1})\rho\ .
\label{qbr}
\eea 
In the case of the algebra $W_3$ we also have
\bea
q^{(3)}_\al = q_\al = \frac{i}{27}(\al_1-\al_2)(2\al_1+\al_2-3b-3b^{-1})(\al_1+2\al_2-3b-3b^{-1})\ .
\label{qa}
\eea
In general, $q^{(s)}$ is Weyl-invariant and homogeneous of degree $s$ as a function of $\al-Q$. The representations which appear in the spectrum of $s\ell_3$ conformal Toda theory have momenta
\bea
\al \in Q+ i(\R \om _1+\R \om_2)\ ,
\label{spec}
\eea
so that $\Delta_\al$ and $q_\al$ are real numbers. 
We will call $R_\al$ the representation with momentum $\al$, and $V_\al(z)$ the corresponding primary vertex operator.
Under a Weyl transformation of the momentum, $V_\al(z)$ is supposed to behave as 
\bea
V_{\al}(z)= R_w(\al)\ V_{Q+w(\al-Q)}(z)\ ,
\label{vrv}
\eea
for some reflection coefficients $R_w(\al)$, and the representation $R_\al$ is unchanged, namely $R_\al = R_{Q+w(\al-Q)}$. On the other hand, the conjugation of $\al$ does not leave the charges invariant, but they transform according to $q^{(s)}_{\al^*}=(-1)^s q^{(s)}_\al$, because $\al^*$ is related to $2Q-\al$ by a Weyl transformation. The conjugation of $\al$ therefore corresponds to the automorphism $W^{(s)}\rar (-1)^s W^{(s)}$ of the algebra $W_n$. (This assumes that the charges $W^{(s\geq 3)}$ are defined so that they are primary operators of dimensions $s$ with respect to $T(z)$; other definitions are in principle possible.) 

A representation is called degenerate if it has one or more null vectors. Let us consider a representation with momentum $\al$. If there is a (positive or negative) root $e$ and two strictly positive integers $r$ and $s$ such that 
\bea
(e,\al-Q) = -rb -sb^{-1}\ ,
\label{ers}
\eea
then our representation has a null vector at level $rs$, which is itself the highest-weight vector of a representation with momentum $\al' = \al + rb e$ (or equivalently $\al''=\al + sb^{-1}e$, which is related to $\al'$ by a Weyl transformation). For any momentum $\al$ let us introduce the set
\bea
E(\al) = \{{\rm roots }\ e\ {\rm such\ that}\ (e,\al-Q)\in -b\N^* -b^{-1}\N^*\}\ .
\eea
By applying a Weyl tranformation to $\al$ we can always assume that $E(\al)$ contains only positive roots. 
A representation $R_\al$ is said to be multiply degenerate if $|E(\al)|\geq 2$, and 
fully degenerate if $E(\al)=\{e>0\}$ so that $|E(\al)|=\frac12 n(n-1)$. The momentum $\al$ of a fully degenerate representation can be written in terms of a pair 
$(\Omega^+,\Omega^-)$ of integral dominant weights of $s\ell_n$ as $\al = -b\Omega^+ - b^{-1} \Omega^-$. In the case of the $W_3$ algebra, we will therefore distinguish three types of degenerate representations: simply degenerate representations with $E(\al)=\{e_1\}$, doubly degenerate representations with $E(\al)=\{e_1,\rho\}$, and fully degenerate representations with $E(\al)=\{e_1,e_2,\rho\}$.

The character of a representation $R_\al$ of the $W_n$ algebra is defined by
\bea
\xi_\al(\tau) \equiv {\rm Tr}_{R_\al}\ e^{2i\pi \tau (L_0-\frac{c}{24})}\ .
\label{xtr}
\eea
This is easily computed in the case of a continuous representation, with the result
\bea
\xi_\al(\tau)= \frac{e^{-i\pi \tau (Q-\al)^2}}{\eta(e^{2i\pi \tau})^{n-1}}\  ,
\label{xat}
\eea
where $\eta$ is the Dedekind eta function.
Let us now consider the case of 
a fully degenerate representation $R_{-b\Omega^+-b^{-1}\Omega^-}$, where $\Omega^\pm$ are integral dominant weights.
The corresponding character is a sum over the Weyl group \cite{bw93},
\bea
\xi_{-b\Omega^+-b^{-1}\Omega^-}(\tau)= \frac{\sum_{w\in W} \e(w)\ e^{-i\pi\tau (b(\rho+\Omega^+)+b^{-1}w(\rho+\Omega^-))^2}}{\eta(e^{2i\pi \tau})^{n-1}}\  .
\label{xw}
\eea
Now we observe that this degenerate character can be expressed in terms of the characters
$\chi_{\Omega^\pm}$ of the two representations of $s\ell_n$ of highest weights $\Omega^\pm$, 
\begin{multline}
\xi_{-b\Omega^+-b^{-1}\Omega^-}(\tau) = \frac{\sqrt{n}}{n!}\int d^{(n-1)}p\ \frac{e^{-\frac{2i\pi}{\tau}\frac12 p^2}}{\eta\left(e^{-\frac{2i\pi}{\tau}}\right)^{n-1}} 
\\
\times \prod_{\pm}\left[ \chi_{\Omega^\pm}(2\pi b^{\pm 1} p)\ \prod_{e>0} 
(e^{\frac12(e,2\pi b^{\pm 1}p)}-e^{-\frac12(e,2\pi b^{\pm 1}p)})\right]\  ,
\label{xoo}
\end{multline}
where the integration measure is defined as $d^{(n-1)}p=\prod_{i=1}^{n-1}dp_i$ with $p=\sum_i p_ie_i$, and
we used the Weyl formula (\ref{cop}).

Notice that the $W_n$ characters $\xi_\al(\tau)$ keep track of the conformal dimensions ($L_0$ eigenvalues) of states, and not of their charges $q^{(s>2)}$. So if $n>2$ they contain much less information than the $s\ell_n$ characters $\chi_\Omega(p)$, which depend on a vector $p$ and not on a single number $\tau$. This will make the modular bootstrap analysis less powerful in theories with $W_{n>2}$ symmetries than in theories with just the Virasoro symmetry.

\subsection{Fusion multiplicity \label{ssmul}}

We will now comment on the fusion product of $W_n$ representations. The fusion product is a generalisation to vertex operator algebras of the tensor product of representations of Lie algebras. So we first comment on the tensor product of $s\ell_n$ representations. We consider generic representations, which are not necessarily finite-dimensional, and even do not necessarily have a highest weight.

The algebra $s\ell_n$ can be represented in terms of differential operators acting on functions of $\frac12 n(n-1)$ ``isospin'' variables. (This is also the number of the creation operators, the operators which generate the highest-weight representations from their highest-weight states.) 
For example, $s\ell_2$ is represented by the operators $D^-=\pp{x},\ D^3=x\pp{x} -j,\ D^+=x^2\pp{x} -2j x$ acting on functions of one isospin variable $x$, where the number $j$ is the spin of the representation. States in a representation of $s\ell_n$ with spin $j$ can be represented as functions $\Psi^j(x)$, where the spin $j$ is a vector with $n-1$ components, and the isospin $x$ is a vector with $\frac12 n(n-1)$ components. We wish to analyse the possible appearances of a representation $R_{j_3}$ in the tensor product $R_{j_1}\otimes R_{j_2}$. Such an appearance implies the existence of a nonzero invariant vector in $R_{j_1}\otimes R_{j_2}\otimes R_{j_3}^*$, where $R_{j_3}^*$ is the contragredient representation. In the representation of $s\ell_n$ in terms of differential operators, an invariant vector in $R_{j_1}\otimes R_{j_2}\otimes R_{j_3}^*$ is represented as a function $\Phi(x_1,x_2,x_3)$ of three isospin vectors, subject to $(\dim s\ell_n) = n^2-1$ equations. If the representations are generic and no more assumptions are made, solutions $\Phi(x_1,x_2,x_3)$ come with the number of parameters
\bea
d_n= 3\frac{n(n-1)}{2} - (n^2-1) = \frac12 (n-1)(n-2)\ .
\label{dn}
\eea
If $n>2$ then $d_n>0$ which implies that $R_{j_3}$ can appear an infinite number of times in $R_{j_1}\otimes R_{j_2}$. If however one of the three representations $R_{j_1}$, $R_{j_2}$ or $R_{j_3}$ is not generic, then extra equations on $\Phi(x_1,x_2,x_3)$ can follow, and the number of parameters may become lower. If the number of parameters is zero, as happens if $n=2$ or one of the involved representations has a highest weight state, then multiplicities must be finite. 

A similar counting of variables, and similar conclusions on fusion multiplicities, hold in the case of the fusion product of $W_n$ representations. This is a consequence of the conformal Ward identities for the three-point correlation functions $\la \prod_{i=1}^3 V_{\al_i}(z_i)\ra$ where the momenta $\al_i$ label $W_n$ representations; such correlation functions are analogous to the invariants $\Phi(x_1,x_2,x_3)$ of our $s\ell_n$ reasoning (although the positions $z_i$ are not analogous to the isospins $x_i$). The fusion multiplicity is the minimum number of correlation functions of descendent operators $\la \prod_{i=1}^3 D_i V_{\al_i}(z_i) \ra$ in terms of which all other such correlation functions can be linearly expressed using the Ward identities. Such identities are obtained by inserting the identity $\oint_\infty \varphi_s(z) W^{(s)}(z) =0$ in a correlation function, where $\oint_\infty$ denotes the integration along a contour which encloses all the positions $z_i$ of the operators, $\varphi_s(z)$ is meromorphic with possible poles at $z=z_i$, and at infinity $|\varphi_s(z)|\leq |z|^{2s-2}$. (We assume $W^{(s)}(z)\underset{z\rar \infty}{\sim} z^{-2s}$, which follows from the $W^{(s)}$-symmetry of the vacuum.) Local Ward identities are obtained for functions $\varphi_s(z)$ which do have poles; the case $\varphi_s(z)=(z-z_i)^{-k}$ (with $k\in\N$) yields the expression of a correlation function involving $W^{(s)}_{1-s-k}V_{\al_i}(z_i)$ in terms of correlation functions with descendents of the type $W^{(s)}_{-p}V_{\al_i}(z_i)$ with $1\leq p\leq s-1$, as can be seen from the operator product $W^{(s)}(z)V_\al(w)$. (See eq. (\ref{wv}) for the case $s=3$.) In the theory with $W_n$ symmetry there are $\frac12 n(n-1)$ modes of the type $W^{(s)}_{-p}$ with $2\leq s\leq n,\ 1\leq p\leq s-1$, for instance the three modes $L_{-1},W_{-1},W_{-2}$ in the case $n=3$, and these modes are analogous to the isospin variables $x$ of the $s\ell_n$ algebra. Global Ward identities are obtained for holomorphic functions $\varphi_s(z)$, that is polynomials of degrees at most $2s-2$. The number of global Ward identities is therefore 
$\sum_{s=2}^n (2s-1)=n^2-1$.

Thus,  the number of modes of the $W^{(s)}$ symmetry generators which cannot be eliminated from the correlation functions of the type $\la \prod_{i=1}^3 D_i V_{\al_i}(z_i) \ra$ using the Ward identities is $d_n$ (\ref{dn}). For example, $d_3=1$ means that in a theory with $W_3$ symmetry all three-point functions can be expressed in terms of the correlation functions $\la V_{\al_1}(z_1)V_{\al_2}(z_2)(W_{-1})^k V_{\al_3}(z_3)\ra$ where $V_{\al_i}$ are primary operators and $k\in \N$. (Instead of $W_{-1}$ we may have written $W_{-2}$, but not $L_{-1}$, because the three global Ward identities from $T(z)$ close among themselves and can be solved.)
Notice that a similar reasoning can be used to predict the number $E$ of independent differential equations obeyed by an $N$-point function of primary operators, some of which may be degenerate and involve a total number $V$ of null vectors. We find $E=n^2-4-N(\frac{n(n-1)}{2}-1)+V$, where we subtract the three equations from global conformal symmetry, as well as the contribution of the $L_{-1}$ generator as it is identified with a derivative. Such a counting of differential equations has previously been used in \cite{rib08b}.

To conclude, infinite fusion multiplicities must appear in all theories with a $W_{n\geq 3}$ symmetry as soon as continuous representations are involved, which will be the case in conformal Toda theories.

\subsection{Lagrangian formulation}

Conformal $s\ell_n$ Toda theory on a Riemann surface without boundary has a Lagrangian formulation. The dynamical fields of the theory form a vector with $n-1$ components
\bea
\phi=\sum_i\phi_i e_i \qquad {\rm so\ that}\qquad\phi_i=(\om_i,\phi)\ ,
\eea
and the Lagrangian is
\bea
L_n = \frac{1}{2\pi} (\p\phi,\bp \phi) + \mu \sum_{i=1}^{n-1} e^{b(e_i,\phi)}\ ,
\label{ln}
\eea
where $\mu$ is the bulk cosmological constant, and the derivatives with respect to the complex coordinates $z,\bz$ are related to derivatives with respect to the real coordinates $x=\Re z,\ y=\Im z$ by $\p=\frac12(\pp{x}-i\pp{y}), \ \bp=\frac12(\pp{x}+i\pp{y})$. After the rescaling $\phi\rar b^{-1}\phi$, the classical equations of motion are 
\bea
\p\bp \phi_i = \pi b^2\mu e^{(e_i,\phi)}\ .
\label{eom}
\eea
The Lagrangian formulation permits the calculation of certain particular correlation functions, and of general correlation functions in certain limits, but not of general correlation function \cite{zz95,fl07c}. For our purposes, we will only make use of the classical equations of motion, and not of functional integrals involving the action $S=\int L_n$. We will actually solve the equations of motion in Section \ref{sted}.

The $W_n$ symmetry of $s\ell_n$ Toda theory manifests itself by the existence of charges $W^{(2)}=T,W^{(3)}\cdots W^{(n)}$ which are classically conserved in the sense that $\bp W^{(s)}=0$. In the case of Liouville theory, this is 
\bea
T=-(\p\phi)^2+\p^2\phi\ .
\eea
In the case of $s\ell_3$ Toda theory, $W^{(3)}=W$ has the ambiguity $W\rar W-\xi \p T$ for $\xi$ an arbitrary number, which we lift by assuming $W(\phi^*)=-W(\phi)$, where $\phi\rar \phi^*$ is the Dynkin diagram automorphism $\phi_1\lrar \phi_2$. We then have 
\bea
T&=&-\tfrac12(\p \phi,\p\phi) + (\rho,\p^2\phi) =-\p\phi_1^2-\p\phi_2^2+\p\phi_1\p\phi_2+\p^2\phi_1+\p^2\phi_2\ ,
\label{tdef}
\\
W&=& \left(\p^3\phi_2-2\p^2\phi_2\p\phi_2-\p^2\phi_2\p\phi_1+2\p\phi_2^2\p\phi_1\right) -(\phi_1\lrar \phi_2)\ ,
\label{wdef}
\eea
where we neglect a possible normalization factor in the definition of $W$. Such classically conserved charges can alternatively be found as classical limits of the corresponding quantum symmetry generators  of the $W_3$ algebra \cite{fz86b}. We use the same notation for the classical charge and the quantum generator; the context should clarify which one we are dealing with. 

The momentum $\al$, which we used as a label for $W_n$ representations (see (\ref{da},\ref{qa})), has a simple interpretation in the Lagrangian formulation. Namely, the classical counterpart of the quantum operator $V_{\al}(z)$ is $e^{(\al,\phi(z))}$.

\zeq\section{Solutions of the Toda equations on a disc \label{sted}}

The classical $s\ell_n$ Toda equations on the Riemann sphere have been solved in \cite{ls79}, see also \cite{af09}. We will now look for solutions on the disc, which are solutions on the sphere respecting certain boundary conditions. We will only consider maximally symmetric boundary conditions, that is conditions of the type $\bar{W}^{(s)}=f(\{W^{(s')}\})$ where $f$ is an automorphism of the $W_n$ algebra. The known automorphisms are the identity, and in the $W_{n\geq 3}$ algebra the automorphism $W^{(s)}\rar (-1)^s W^{(s)}$. There will therefore be two possible types of boundary conditions if $n\geq 3$, and only one if $n=2$.
We will study the cases of the $W_2$ and $W_3$ algebras. The sphere will be identified with the complex plane and parametrized by the coordinates $z,\bz$, and the disc will be identified with the upper half-plane $\{\Im z>0\}$. 

In this Section, we will study the solutions of the Toda equations and in particular their invariants, which we call the boundary parameters. The question whether our boundary conditions follow from boundary actions is postponed to Appendix \ref{secac}. 
We will consider the classical Toda equations (\ref{eom}) with the value 
\bea
\mu=-\frac{1}{\pi b^2}
\label{muval}
\eea
for the cosmological constant. The choice of a negative value for $\mu$ will allow real, globally defined, regular solutions to exist.

\subsection{Case of Liouville theory \label{cali}}

In order to solve the Liouville equation $\p\bp \phi= -e^{2\phi}$ together with the boundary condition $T=\bar{T}$ where $T=-(\p\phi)^2+\p^2\phi$, we introduce the variable $X=e^{-\phi}$ which is such that $T=-\frac{\p^2 X}{X}$ and the Liouville equation amounts to $\Delta_2(X)\equiv X\p\bp X-\p X\bp X=1$. The solutions of this equation are of the form
\bea
X=\sum_{i=1}^2 a_i(z)b_i(\bz) \scs {\rm Wr}[a_1,a_2]={\rm Wr}[b_1,b_2]=1\ ,
\eea 
where ${\rm Wr}[a_1,a_2]=a_1a_2'-a_2a_1'$ is the Wronskian. (By setting ${\rm Wr}[a_1,a_2]={\rm Wr}[b_1,b_2]=1$ we have eliminated the ambiguity $a_i\rar \xi a_i,\ b_i\rar \xi^{-1} b_i$.) The stress-energy tensor $T$ associated with such a solution is $T=-\frac{a_1''}{a_1}=-\frac{a_2''}{a_2}$. The condition $T=\bar{T}$ is solved by assuming 
\bea
a_i(z)=\sum_{j=1}^2 N_{ij} b_j(z) \scs \det N=1 \ .
\eea 
The condition that $X$ be real and positive will now be solved by assuming that $b(z)=\bar{b}(z)$ is real, and that the constant matrix $N$ is Hermitian and positive. To summarize, our solutions are 
\bea
X=\sum_{i,j=1}^2 \overline{b_i(z)} N_{ij} b_j(z) \scs {\rm Wr}[b_1,b_2]=1\ ,\ \det N=1,\ N>0\ .
\label{xsol}
\eea
There remain some ambiguities in the solutions, because different choices for $N_{ij},b_i(z)$ can lead to the same  $X$. In particular, the following action of $\SLR$ leaves $X$ invariant:
\bea
\left\{\begin{array}{l} b\rar \Lambda^{-1} b \\ N\rar \Lambda^T N \Lambda \end{array}\right. \scs \Lambda\in \SLR\ ,
\label{linv}
\eea
where $\Lambda^T$ denotes the transpose of the matrix $\Lambda$. 

Let us define a boundary parameter $\lambda_L$ associated to a given solution. We assume this parameter to be a $z$-independent function of the solution $X$. Independence from $z$ implies being a function of the matrix $N$. Being a function of $X$ implies being invariant under the action (\ref{linv}) of $\SLR$. The only such invariant function of $N$ is 
\bea
\lambda_L \equiv \frac{1}{2i}\Tr NP \scs P\equiv \bsm  0 & -1 \\ 1 & 0 \esm \ .
\label{trnp}
\eea
Notice that the matrix $P$ obeys $\forall \Lambda \in \SLR,\ \Lambda P\Lambda^T=P$. The role of $\lambda_L$ as a boundary parameter can be demonstrated by rewriting the boundary condition in terms of the field $X=e^{-\phi}$. At the boundary $z=\bz$ we find:
\bea
(\p-\bp)X= 2i\lambda_L \scs (\p -\bp)\phi=-2i\lambda_L e^{\phi}\ .
\eea
This implies that the boundary conditions could be derived by adding a boundary term $\int L_2^{bdy}$ to the action, with 
\bea
L_2^{bdy}= \lambda_L e^\phi\ ,
\label{ltb}
\eea
and $\lambda_L$ would be the boundary cosmological constant. 

We conclude this Subsection with a remark.
Given the solution of the Liouville equation, it is easy to write the B\"acklund transformation from Liouville theory to a free field theory. The free field can be defined as $\psi = -\log \frac{\sum_i u_ia_i(z)}{\sum_i v_ib_i(\bz)}$ where $(u_1,u_2)$ and $(v_1,v_2)$ are constant vectors, and the stress-energy tensors are $T=-(\p\psi)^2+\p^2\psi,\ \bar{T}=-(\bp\psi)^2-\bp^2\psi$. The field $\psi$ obeys the free equations of motion $\p\bp\psi=0$, as well as Dirichlet boundary conditions $(\p+\bp)\psi=0$, and the value of $\psi$ at the boundary is related to $\lambda_L$. 

\subsection{Case of $s\ell_3$ Toda theory}

Let us solve the $s\ell_3$ Toda equations $\bla \p\bp \phi_1 = -e^{2\phi_1-\phi_2} \\ \p\bp \phi_2= -e^{2\phi_2-\phi_1} \ela$. In terms of $X_i=e^{-\phi_i}$, the $s\ell_3$ Toda equations amount to $\bla \Delta_2(X_1)=X_2 \\ \Delta_2(X_2)=X_1\ela$, where $\Delta_2(X)$ was defined in the previous Subsection, and is such that $\Delta_2(\Delta_2(X))=X\det\bsm X & \p X & \p\p X \\ \bp X & \p\bp X & \p\p\bp X \\ \bp\bp X & \p \bp\bp X & \p\p\bp\bp X \esm$. The solutions of the $s\ell_3$ Toda equations are 
\bea
\bla
X_1=\sum_{i=1}^3 a_i(z) b_i(\bz)\\ X_2=\sum_{i<j} {\rm Wr}[a_i,a_j](z) {\rm Wr}[b_i,b_j](\bz)
\ela \scs {\rm Wr}[a_1,a_2,a_3]={\rm Wr}[b_1,b_2,b_3]=1\ ,
\eea
where ${\rm Wr}[a_1,a_2,a_3]=\e_{ijk} a_ia'_ja''_k$ is the cubic Wronskian. In this solution, the Dynkin diagram automorphism $\phi_1\lrar \phi_2$ manifests itself as $\bla a_i\lrar \frac12\e_{ijk}{\rm Wr}[a_j,a_k] \\ b_i\lrar \frac12\e_{ijk}{\rm Wr}[b_j,b_k] \ela$. In these formulas we used the fully antisymmetric tensor $\e_{ijk}$ such that $\e_{123}=1$. 

Let us rewrite the symmetry charges $T$ (\ref{tdef}) and $W$ (\ref{wdef}) in terms of the variables $X_1,X_2$ or $a_i,b_j$: 
\bea
T&=& -\frac{\p^2X_1}{X_1}-\frac{\p^2 X_2}{X_2} +\frac{\p X_1}{X_1}\frac{\p X_2}{X_2} = \frac{\p^3 X_1 \bp X_1-X_1\p^3\bp X_1 }{X_2}=\frac{a_3 a_1'''-a_1a_3'''}{{\rm Wr}[a_1,a_3]}\ ,
\label{tfield}
\\
W&=& T\left(\frac{\p X_2}{X_2}-\frac{\p X_1}{X_1}\right) -\frac{\p^3 X_1}{X_1}+\frac{\p^3 X_2}{X_2}=-T'+2\frac{a_1'a_3'''-a_3'a_1'''}{{\rm Wr}[a_1,a_3]}\ ,
\label{wfield}
\eea
where we could use any pair of functions $a_i$ instead of $(a_1,a_3)$, and the result would not change due to the identity ${\rm Wr}[a_1,a_2,a_3]'=0$. The antiholomorphic charges $\bar{T},\bar{W}$ are similarly written in terms of $b_i$.

We now consider $s\ell_3$ Toda theory on the half-plane, and the possible boundary conditions on the real line. We must impose $T=\bar{T}$ for conformal symmetry to be preserved. For the spin $3$ current $W$ we have the two choices $W=\pm \bar{W}$, where the minus sign corresponds to using the nontrivial automorphism of the $W_3$ algebra.

\subsection{Boundary condition $W-\bar{W}=0$ \label{ssm}}

Now that we wrote the solutions of the bulk equations of motion in terms of the functions $a_i,b_i$, let us write boundary conditions for these functions. The conditions $T=\bar{T},W=\bar{W}$ are obeyed provided we assume
\bea
a_i=\sum_{j=1}^3 N_{ij} b_j \scs \det N=1 \ .
\label{anb}
\eea
This implies $\e_{ijk}{\rm Wr}[a_j,a_k] = (N^{-1T})_{ii'} \e_{i'j'k'}{\rm Wr}[b_{j'},b_{k'}]$, where $N^{-1T}$ is the inverse of the transpose of $N$. 
Furthermore, in order for $X_1,X_2$ to be positive, we assume $b_i=\bar{b}_i$ and that $N$ is a positive Hermitian matrix. To summarize,
\bea
\bla X_1=\sum_{i,j=1}^3 \overline{b_i(z)} N_{ij} b_j(z)
 \\ X_2 = \sum_{i,j=1}^3 \overline{w_i(z)} (N^{-1T})_{ij} w_j(z) \ela
 \scs {\rm Wr}[b_1,b_2,b_3]=1\ ,\ \det N=1,\ N>0\ ,
\label{xmsol}
\eea
where we introduced the notation $w_i= \e_{ijk}b_j b'_k$ for the quadratic Wronskians, whose quadratic Wronskians are themselves ${\rm Wr}[w_i,w_j]=\e_{ijk}b_k$. 

As in the case of Liouville theory, our parametrization of the solutions is ambiguous, because $X_1,X_2$ are invariant under the action of an $SL(3,\R)$ symmetry group, namely
\bea
\left\{\begin{array}{l} b\rar \Lambda^{-1} b \\ w\rar \Lambda^T w \\ N\rar \Lambda^T N \Lambda \end{array}\right. \scs \Lambda\in SL(3,\R)\ .
\label{tminv}
\eea
The boundary parameters are the $SL(3,\R)$-invariant functions of $N$. Such invariants can be constructed as $\Tr(N^TN^{-1})^m,\ m=1,2\cdots$. Given $\det N=1$, all such invariants are functions of 
\bea
\lambda_0 \equiv \det \tfrac12(N+N^T)\ .
\label{ldn}
\eea
In the $s\ell_n$ case the number of invariants is the integer part of $\frac{n}{2}$.

We may wish to express the boundary conditions in terms of the fields $X_i=e^{-\phi_i}$. To this end, we may compute at the boundary
\bea
\bla (\p-\bp)X_1=\tfrac12 \e_{ijk} N_{ij} w_k 
\\ (\p-\bp)X_2 = \tfrac12 \e_{ijk} N^{-1T}_{ij} b_k \ela\ .
\label{dxm}
\eea
The right hand sides of these expressions are in general not functions of $X_1,X_2$. There is an exception in the special case when $N^T=N$, which corresponds to the free boundary conditions $(\p-\bp)X_i=0$, in which case the boundary parameter is $\lambda_0=1$. Another exception occurs when $N_{ij}=U_iU_j+\e_{ijk}A_k$, with $A_iU_i=1$ so that $\det N=1$. (We drop the assumption that $N$ be hermitian.) In this case the boundary parameter is $\lambda_0=-1$. 
Noticing $N^{-1T}=A_iA_j+\e_{ijk}U_k$, we find $\bla (\p-\bp)X_1=A_k w_k =\sqrt{X_2}
\\ (\p-\bp)X_2 = U_k b_k =\sqrt{X_1} \ela$. Such boundary conditions derive from the boundary Lagrangian
\bea
L_3^{bdy} = \frac{1}{2i}\left(e^{\phi_1-\frac12 \phi_2} + e^{\phi_2-\frac12\phi_1} \right)\ .
\label{lz}
\eea
One may be tempted to generalize this Lagrangian into $L_3^{bdy}=\nu_1 e^{\phi_1-\frac12\phi_2} +\nu_2 e^{\phi_2-\frac12\phi_1}$, which would depend on two boundary parameters $\nu_1,\nu_2$. However, it turns out that for general values of $\nu_1,\nu_2$ the $W_3$ symmetry would then be broken, in the sense that the boundary condition $W=\bar{W}$ would not be obeyed. Only the values of $\nu_1,\nu_2$ which we wrote in eq. (\ref{lz}) are therefore permitted.

\subsection{Boundary condition $W+\bar{W}=0$ \label{ssp}}

Boundary conditions for the functions $a_i,b_i$ which imply $W+\bar{W}=0$ are now
\bea
a_i=\sum_{j=1}^3 N_{ij} w_j \scs \det N=1 \ .
\label{anw}
\eea
This can be deduced from the case $W-\bar{W}=0$ by using the Dynkin diagram automorphism, 
which exchanges the functions $b_i$ with their Wronskians $w_i=\e_{ijk}b_jb'_k$. It is however not clear how to guarantee the positivity of $X_1,X_2$. We refrain from making further assumptions on the matrix $N$, and we write the solutions of the classical Toda equations as
\bea
\bla X_1=\sum_{i,j=1}^3 b_i(\bz) N_{ij} w_j(z)
 \\ X_2 = \sum_{i,j=1}^3 w_i(\bz) N^{-1T}_{ij} b_j(z) \ela
 \scs {\rm Wr}[b_1,b_2,b_3]=1\ ,\ \det N=1\ .
\label{xpsol}
\eea
As we do not impose reality conditions on $b_i$ and $N$ we 
find that $X_1,X_2$ are invariant under the action of an $SL(3,\C)$ group of symmetries, instead of an $SL(3,\R)$ group in the case $W-\bar{W}=0$:
\bea
\left\{\begin{array}{l} b\rar \Lambda^{-1} b \\ w\rar \Lambda^T w \\ N\rar \Lambda^T N \Lambda^{-1T}\end{array}\right. \scs \Lambda\in SL(3,\C)\ .
\label{tpinv}
\eea
The group acts by conjugation on the matrix $N$, and there are two invariants, which we interpret as boundary parameters:
\bea
\lambda_1 = \Tr N \scs \lambda_2 = 
\Tr N^{-1}\ . 
\label{ltn}
\eea
In the $s\ell_n$ case the number of invariants is of course $n-1$. This corresponds to the number of conserved charges $W^{(2)}\cdots W^{(n)}$. This already suggests that the boundary condition $W+\bar{W}=0$ realizes Cardy's ideas on the correspondence between representations of the symmetry algebra and boundary parameters. We will demonstrate this further in our conformal bootstrap analysis in Section \ref{secbo}.

An interesting case happens when the matrix $N$ obeys a second-order polynomial equation, that is when two of its eigenvalues coincide. Then we have $N^{-1}=uN+vI$ where $I$ is the identity matrix and $u,v$ are two complex numbers, and it follows from eq. (\ref{xpsol}) that $\phi_1-\phi_2$ obeys Dirichlet boundary conditions and $\phi_1+\phi_2$ obeys Neumann boundary conditions,
\bea
\phi_1-\phi_2=c \scs (\p-\bp)(\phi_1+\phi_2)=0\ ,
\label{dnc}
\eea
where $c$ is an arbitrary constant. 
These conditions can be derived from the $s\ell_3$ Toda action with no boundary terms. Due to the Dirichlet condition, the spacetime interpretation of this case is a one-dimensional D-brane, whereas the other cases describe two-dimensional D-branes. The one-dimensional D-branes extend along the direction of the Weyl vector $\rho$, which is consistent with the existence of a linear dilaton in that direction. (This linear dilaton can be seen in the expression for the stress-energy tensor $T$ (\ref{tdef}).) 

We conclude this Subsection with a remark.
Given the solution of the Toda equations, it is easy to write the B\"acklund transformation from Toda field theory to a free field theory. The free fields can be defined as $\psi_i=-\log \frac{\sum_j u_{ij} a_j(z)}{\sum_j v_{ij} w_j(\bz)}$ where $u_{ij}$ and $v_{ij}$ are constant matrices. The free fields obey free equations of motion $\p\bp\psi_i=0$, and Dirichlet boundary conditions for $\psi_i$ imply relations of the type $a_i=\sum_j N_{ij}w_j$ (\ref{anw}) at the boundary. Thus, we can interpret the boundary parameters $\lambda_1,\lambda_2$ encoded in the matrix $N$ as the boundary values of the free fields. (See Appendix \ref{secmi} for more details on the B\"acklund transformation.)

\zeq\section{Conformal bootstrap study of $s\ell_3$ Toda theory with $W+\bar{W}=0$ \label{secbo}}

The conformal bootstrap method is the systematic exploitation of symmetry and consistency constraints on correlation functions in two-dimensional conformal field theories \cite{bpz84}. We will apply this method to the correlation functions of $W_3$ primary operators $V_\al(z)$. Such operators are defined up to normalizations by their operator products (\ref{tv}) and (\ref{wv}) with the symmetry generators $T(z),W(z),\bar{T}(\bz),\bar{W}(\bz)$. 
\footnote{
Due to the existence of the $\Z_2$ automorphism of the $W_3$ algebra, there are two possible definitions of $\bar{W}$, which differ by a sign. Our definition is such that the vertex operators $V_\al(z)$ have the same charge $q_\al$ with respect to $W$ and $\bar{W}$. Then the untwisted (or Cardy) 
boundary condition is $W=-\bar{W}$. We thank Gor Sarkissian for correspondence which led to this clarification.
}
In order to define correlation functions on the upper half-plane, the properties of the boundary must be characterized. This involves first of all imposing
boundary conditions for the symmetry generators $T(z),W(z)$. Moreover, we saw in Section \ref{sted} that for each of the two boundary conditions $\bla T=\bar{T} \\ W=\pm \bar{W}\ela $ there exist families of possible D-branes, parametrized by $\lambda_0$ or $\lambda_1,\lambda_2$. These parameters appeared in the classical analysis of the boundary conditions for the basic Toda fields $\phi_i$, but such fields are not present in the conformal bootstrap formalism. Nevertheless, the equivalents of $\lambda_0$ or $\lambda_1,\lambda_2$ will appear when we will parametrize the solutions of the conformal bootstrap equations; a given solution will be called a D-brane or boundary state. 

In the case of Liouville theory there is only one possible boundary condition $T=\bar{T}$, and there exist two types of D-branes: the continuous D-branes \cite{fzz00,tes00}, with a continuous parameter, and the discrete D-branes \cite{zz01}, which are parametrized by two integers. These two types of D-branes are associated to the two types of representations of the Virasoro algebra: the continuous and discrete representations. By analogy, we expect that in conformal $s\ell_n$ Toda theory there exists a hierarchy of D-branes, which would correspond to the hierarchy of representations of the $W_n$ algebra which we discussed in Subsection \ref{ssrep}. The dimension of a D-brane would be $n-1-k$, where $k$ is the number of algebraically independent null vectors in the corresponding representation. In the case of $s\ell_3$ Toda theory, we would have three types of D-branes: two-dimensional continuous D-branes, one-dimensional ``simply degenerate'' D-branes, and zero-dimensional discrete or fully-degenerate D-branes. We will see that these expectations are fulfilled when the boundary condition is $W+\bar{W}=0$. 

An important difference between the two boundary conditions $W=\pm \bar{W}$ manifests itself when analyzing the consequences of the $W_3$ symmetry on the correlation functions. We introduced the Ward identities which follow from the $W_3$ symmetry in Subsection \ref{ssmul}, let us now sketch how such identities constrain the correlation function of $N$ operators $V_{\al_i}(z_i)$ (with $\Im z_i>0$) in the presence of a boundary at $z=\bz$. It turns out that the Ward identities for such an $N$-point function are identical to the Ward identities for a $2N$-point function in the absence of a boundary, where the extra $N$ operators are ``reflected'' operators located at $\bz_i$. The reflected operators are $V_{\al_i}(\bz_i)$ if the boundary condition is $W-\bar{W}=0$, and $V_{\al_i^*}(\bz_i)$ if the boundary condition is $W+\bar{W}=0$. As far as the Ward identities are concerned, we thus have the relations
\bea
\la V_\al(z) \ra_{W-\bar{W}=0} \sim \la V_\al(z) V_\al(\bz) \ra \ ,
\label{vmvv}
\\
\la V_\al(z) \ra_{W+\bar{W}=0} \sim \la V_\al(z) V_{\al^*}(\bz) \ra\ .
\label{vpvv}
\eea
The Ward identities for a bulk two-point function $\la V_{\al_1}(z_1)V_{\al_2}(z_2)\ra$ are known to imply that it vanishes unless $\Delta_{\al_1}=\Delta_{\al_2}$ and $q_{\al_1}+q_{\al_2}=0$.
Now eqs. (\ref{da}) and (\ref{qa}) imply $\Delta_{\al}=\Delta_{\al^*}$ and $q_{\al}=-q_{\al^*}$. Therefore, while $\la V_\al(z) \ra_{W+\bar{W}=0}$ may be nonzero for all values of $\al$, $\la V_\al(z) \ra_{W-\bar{W}=0}$ must vanish unless $q_\al=0$. This restricts the momentum $\al$ to a one-dimensional space, which may be related to the fact that there is only one boundary parameter $\lambda_0$ in the case $W-\bar{W}=0$. We will however not analyze this case further, and instead concentrate on the case $W+\bar{W}=0$ from now on.

\subsection{Continuous D-branes}

Due to conformal symmetry, a one-point function on the upper half-plane must take the form
\bea
\la V_\al(z) \ra = \frac{U(\al)}{|z-\bz|^{2\Delta_\al}}\ ,
\label{vu}
\eea 
where $U(\al)$ is the bulk one-point structure constant, which we now want to determine. We will find  constraints on $U(\al)$ by considering the two-point function $\la V_{-b\om_1}(y)V_\al(z) \ra$,
which can be factorized in two possible ways:
\bea
\la V_{-b\om_1}(y)V_\al(z) \ra &=& \sum_{h\in H_{\om_1}} C_h(\al) U(\al-bh) {\cal G}_h(\al|y,z)\ ,
\label{vvcug}
\\
&=& \sum_{j} R_j S_j(\al) {\cal F}_j(\al|y,z)\ .
\label{vvbbf}
\eea
Let us 
explain these formulas. The first formula follows from the operator product expansion 
\bea
V_{-b\om_1}V_\al &\rar& \sum_{h\in H_{\om_1}} C_h(\al) V_{\al-bh}\ .
\eea 
This OPE is a sum of three terms labelled by the set $H_{\om_1}=\{\om_1,\om_2-\om_1,-\om_2\}$ of the weights of the fundamental representation of $s\ell_3$; this is analogous to the tensor product of $s\ell_3$ representations $R_{\om_1}\otimes R_\Omega = \sum_{h\in H_{\om_1}} R_{\Omega+h}$. We choose to study the correlation function $\la V_{-b\om_1}(y)V_\al(z) \ra$ precisely because the fully degenerate operator $V_{-b\om_1}$ has such simple OPEs; we could in principle use arbitrary operators instead, but the resulting constraints on $U(\al)$ could not necessarily be written explicitly. The OPE coefficients $C_h(\al)$ are \cite{fl07c}
\bea
C_{\om_1}(\al) &=& 1 \ ,
\label{co}
\\
C_{\om_2-\om_1}(\al) &=& -\frac{\pi \mu}{\g(-b^2)} \frac{\g(b(e_1,\al-Q))}{\g(b(e_1,\al))} \ ,
\label{coo}
\\
C_{-\om_2}(\al) &=& \left(\frac{\pi \mu}{\g(-b^2)}\right)^2 \frac{\g(b(e_2,\al-Q))}{\g(b(e_2,\al))} \frac{\g(b(\rho,\al-Q))}{\g(b(\rho,\al))}\ ,
\label{cmo}
\eea
where we recall that $\mu$ is the bulk cosmological constant, $b$ parametrizes the central charge, and $Q=(b+b^{-1})\rho$ (see Subsection \ref{ssrep}). We also introduce the function $\g(x)=\frac{\G(x)}{\G(1-x)}$ where $\G(x)$ is Euler's Gamma function. The last factor in eq. (\ref{vvcug}) is the conformal block
${\cal G}_h(\al|y,z)$. From our remark that $N$-point functions on the upper half-plane are equivalent to $2N$-point functions on the plane as far as Ward identities are concerned, it follows that ${\cal G}_h(\al|y,z)$ coincides with a bulk four-point $t$-channel conformal block,
\footnote{
In our notation for conformal blocks, all external legs are ``incoming''. ``Incoming'' and ``outgoing'' legs are related by the conjugation $\al\rar \al^*$ of the momentum.
}
\bea
{\cal G}_h(\al|y,z) = 
\psset{unit=.36cm}
\pspicture[](-7,-4)(5,4)
\blockverh{V_{-b\om_1}(y)}{V_{-b\om_2}(\bar{y})}{V_{\al^*}(\bz)}{V_\al(z)}{\al-bh}
\endpspicture
\eea
Explicit expressions for such conformal blocks can be deduced from \cite{fl07c}, where more general conformal blocks were computed. Up to simple common prefactors, the three conformal blocks are
\bea
{\cal G}_h(\al|y,z) \propto \left|\tfrac{y-z}{y-\bz}\right|^{2b(h,\al-Q)}\! \tft{-b^2,}{b(e_h,\al-Q)-b^2,}{b(e'_h,\al-Q)-b^2}{b(e_h,\al-Q)+1,}{b(e'_h,\al-Q)+1}{\left|\tfrac{y-z}{y-\bz}\right|^2}\ ,
\label{ghf}
\eea
where for a given weight $h\in H_{\om_1}$ we call $e_h,e'_h$ the two roots such that $(h,e_h)=(h,e'_h)=1$, for instance $e_{-\om_2}=-e_2,\ e'_{-\om_2}=-\rho$. Similarly, the quantity ${\cal F}_j(\al|y,z)$ in eq. (\ref{vvbbf}) is an $s$-channel conformal block,
\bea
{\cal F}_j(\al|y,z) = 
\psset{unit=.36cm}
\pspicture[](-8,-3)(5,2.8)
\blockhorh{V_{-b\om_1}(y)}{V_{-b\om_2}(\bar{y})}{V_{\al^*}(\bz)}{V_\al(z)}{}{}{j}\ 
\endpspicture
\eea
where $j$ labels the operator which propagates in the $s$-channel. There is a subtlety here: the $t$-channel analysis predicts the existence of three independent blocks, but only two primary operators can appear in the OPE $V_{-b\om_1}V_{-b\om_2}$, namely $V_0$ and $V_{-b\rho}$. The point is that a descendent of $V_{-b\rho}$ can appear independently of $V_{-b\rho}$ itself; we will label as $j=\rho'$ the corresponding $s$-channel operator. In the presence of a boundary, we thus have the bulk-boundary OPE
\bea
V_{-b\om_1} \rar \sum_{j\in\{0,\rho,\rho'\}} R_j B_{-bj}\ ,
\label{vrb}
\eea
where the coefficients $R_j$ are unknown functions of $b$, and $B_{-bj}$ are boundary operators, with the convention that $B_{-b\rho'}$ is some descendent of $B_{-b\rho}$. We will not dwell longer on this subtlety, as we are presently only interested in the $s$-channel operator $j=0$. In this case, the bulk-boundary structure constant $S_j(\al)$ reduces to
\bea
S_0(\al)=U(\al)\ .
\eea
We can now obtain an equation for $U(\al)$ from the equality of (\ref{vvcug}) with (\ref{vvbbf}) by using the fusion tranformation
\bea
{\cal G}_h(\al|y,z) = \sum_j F_{h,j}(\al) {\cal F}_j(\al|y,z)\ ,
\label{gff}
\eea
and extracting the ${\cal F}_0$ term in eq. (\ref{vvcug}). The result is
\bea
\sum_{h\in H_{\om_1}} C_h(\al) F_{h,0}(\al) U(\al-bh) = R_0 U(\al)\ .
\label{cfuru}
\eea
In order to make this equation explicit, let us compute the fusing matrix elements $F_{h,0}(\al)$ defined in eq. (\ref{gff}). Determining $F_{h,j}(\al)$ can be done by taking the limit $\Im z\rar 0$ in that equation. 
By definition the blocks ${\cal F}_j(\al|y,z)$ are power-like functions of $\Im z$ in the limit $\Im z\rar 0$, and we have
\bea
{\cal G}_h(\al|y,z) \underset{\Im z\rar 0}{\sim} F_{h,\rho}(\al) + 
\frac{\Im y \Im z }{|y-\bz|^2} F_{h,\rho'}(\al) + \left(\frac{\Im y \Im z }{|y-\bz|^2}\right)^{2+3b^2} F_{h,0}(\al)\ .
\eea
As the block ${\cal G}_h(\al|y,z)$ is a $_3F_2$ hypergeometric function (\ref{ghf}), let us study such functions, starting with their integral representation 
\begin{multline}
\tft{A_1}{A_2}{A_3}{B_1}{B_2}{z} =\frac{\G(B_2)}{\G(A_3)\G(B_2-A_3)} \int_0^1 dt\ 
t^{A_3-1}(1-t)^{B_2-A_3-1}\ 
\\ \times
{}_2F_1(A_1,A_2,B_1,tz)\ .
\end{multline}
We wish to study the $z\rar 1$ limit, where the critical exponents are $0$, $1$ and $B_1+B_2-A_1-A_2-A_3$, and to focus on the last one of those three exponents. Consider the region $t\rar 1,\ z\rar 1$, where we can use the approximation $_2F_1(A_1,A_2,B_1,tz) \underset{tz\rar 1}{\sim} (1-tz)^{B_1-A_1-A_2} \frac{\G(B_1) \G(A_1+A_2-B_1)}{\G(A_1)\G(A_2)}$. Then we obtain the term with critical exponent $B_1+B_2-A_1-A_2-A_3$,
\begin{multline}
\tft{A_1}{A_2}{A_3}{B_1}{B_2}{z} \underset{z\rar 1}{\supset}\frac{\G(B_1)\G(B_2) \G(A_1+A_2+A_3-B_1-B_2)}{\G(A_1) \G(A_2)\G(A_3)} 
 \\
\times  (1-z)^{B_1+B_2-A_1-A_2-A_3} \ .
\label{tftlim}
\end{multline}
This term may or may not be the leading term of $\tft{A_1}{A_2}{A_3}{B_1}{B_2}{z}$ in the limit $z\rar 1$, depending on the values of $A_i,B_j$. In the case of ${\cal G}_h(\al|y,z)$ (\ref{ghf}), it is actually subleading, as we assume $b>0$ and therefore $B_1+B_2-A_1-A_2-A_3=2+3b^2>2$.
What we are interested in is the coefficient of $(1-z)^{2+3b^2}$, which is 
\bea
F_{h,0}(\al) = \frac{\G(-2-3b^2)}{\G(-b^2)} \frac{\G(b(e_h,\al-Q)+1)}{\G(b(e_h,\al-Q)-b^2)} \frac{\G(b(e'_h,\al-Q)+1)}{\G(b(e'_h,\al-Q)-b^2)}\ .
\label{fhza}
\eea
Combining this formula with the formulas for $C_h(\al)$, eq. (\ref{co})-(\ref{cmo}), we obtain
\bea
F_{h,0}(\al) C_h(\al) = \frac{\G(-2-3b^2)}{\G(-b^2)} \left[-\frac{\pi \mu}{\g(-b^2)}\right] \frac{A(\al-bh)}{A(\al)}\ ,
\label{fc}
\eea
where we introduced the function 
\bea
A(\al) \equiv \left[\pi \mu\g(b^2)\right]^{\frac{(\rho,\al-Q)}{b}}  \prod_{e>0} \G(b(e,\al-Q))^{-1}\G(1+b^{-1}(e,\al-Q))^{-1}\ .
\label{aal}
\eea
This function already appeared in conformal Toda theory \cite{fat01}, as a building block for the reflection coefficient $R_w(\al)=\frac{A(Q+w(\al-Q))}{A(\al)}$ defined in eq. (\ref{vrv}). From that equation and the definition (\ref{vu}) of $U(\al)$, it follows that $A(\al)U(\al)$ must be invariant under the reflections $\al\rar Q+w(\al-Q)$. The equation for $U(\al)$ (\ref{cfuru}) can now be rewritten as
\bea
R_0 A(\al)U(\al) =\sum_{h\in H_{\om_1}}\ A(\al-bh)U(\al-bh)\ ,
\label{rau}
\eea
where $R_0$ is still an unknown function of $b$, in which we actually absorbed the $\al$-independent prefactors of $F_{h,0}(\al) C_h(\al)$ in eq. (\ref{fc}). Three more equations for $A(\al)U(\al)$ can similarly be obtained, by replacing the fully degenerate operator $V_{-b\om_1}$ in eqs. (\ref{vvcug}) and (\ref{vvbbf}) with one of the similar operators $V_{-b\om_2},V_{-b^{-1}\om_1}$ or $V_{-b^{-1}\om_2}$. The resulting equations for $A(\al)U(\al)$ are obtained from eq. (\ref{rau}) by replacing $H_{\om_1}$ with $H_{\om_2}$ and/or $b$ by $b^{-1}$. The coefficient $R_0$ can also change, and we rename it $\lambda_{i,\pm}$ depending on the case. So we obtain the four equations
\bea
\lambda_{i,\pm} A(\al)U(\al) = \sum_{h\in H_{\om_i}}\ A(\al-b^{\pm 1}h) U(\al-b^{\pm 1}h)\ .
\label{lau}
\eea
The smooth, reflection-invariant solutions of these equations are $A(\al)U(\al)=\sum_{w\in W} e^{(w(s),\al-Q)}$ where $W$ is the Weyl group and the arbitary vector $s$ is the boundary parameter. The coefficients $\lambda_{i,\pm}$ might be called the boundary cosmological constants, and their values are 
\bea
\lambda_{i,\pm} = \chi_{\om_i}(-b^{\pm 1} s)\ ,
\label{lch}
\eea 
where we recall that the fundamental character is $\chi_{\om_1}(p)=e^{(\om_1,p)}+e^{(\om_2-\om_1,p)}+e^{-(\om_2,p)}$. The full formula for the solution $U_s(\al)$ is 
\bea
\boxed{U_{s}(\al) =   \left[\pi \mu\g(b^2)\right]^{\frac{(\rho,Q-\al)}{b}}  \prod_{e>0} \G(b(e,\al-Q))\G(1+b^{-1}(e,\al-Q))  \sum_{w\in W} e^{(w(s),\al-Q)}}\ .
\label{opt}
\eea
Our equations (\ref{lau}) being linear, this formula holds up to an $\al$-independent factor. We will say that $U_s(\al)$ defines a continuous D-brane when $s$ is such that $U_s(\al)$ does not diverge exponentially in the limit of large momentum $|\al-Q|\rar \infty$. As the 
operators in the spectrum of $s\ell_n$ Toda theory have purely imaginary values of $\al-Q$ (see eq. (\ref{spec})), the continuous D-branes must have real values of $s$. 

Notice that we wrote the one-point structure constant $U_s(\al)$ in a form which makes sense in $s\ell_n$ Toda theory for arbitrary $n$, and even in conformal Toda theories based on arbitrary simply-laced Lie algebras. We conjecture that this result, and most of the results in the rest of this Section, are valid in the general case and not only in $s\ell_3$ Toda theory.

\subsection{Degenerate D-branes}

In the previous Subsection we found continuous D-branes, whose parameter space has the same dimension as the space of continuous representations. According to the classical analysis of Subsection \ref{ssp}, such D-branes should be interpreted as covering the two-dimensional Toda space whose coordinates are $\phi_1,\phi_2$, because the boundary conditions for the fields $\phi_1,\phi_2$ are of the Neumann type.
We will now investigate 
degenerate D-branes, whose dimensions and parameter spaces should be smaller. 
We will argue in Subsection \ref{ssmo} that the dimension of a D-brane is related to the divergence of its one-point function in the limit $\al-Q\rar 0$: the higher the dimension, the more severe the divergence. So let us
look for solutions $U(\al)$ to the equations (\ref{lau}) whose divergences at $\al-Q\rar 0$ would be less severe than the divergence of $U_s(\al)$ (\ref{opt}). Cancelling some of the divergences from the poles of the $\G(b(e,\al-Q))$ factor can be achieved by taking a linear combination of several solutions $U_s(\al)$ with different values of $s$. However, the resulting combination will still be a solution of eq. (\ref{lau}) only provided the four parameters $\lambda_{i,\pm}$ are the same for all the involved values of $s$.

In order to find two different values $s,s'$ which have the same parameters $\lambda_{i,\pm}$, we make two observations: first, the parameters $\lambda_{i,\pm}=\chi_{\om_i}(-bs)$ (\ref{lch}) are Weyl-invariant, second,
they are invariant under shifts $s\rar s+2\pi i b^{\mp 1}(\Z e_1+\Z e_2)$. Thus, there must exist two elements $w_\pm$ of the Weyl group such that $s-w_\pm (s')\in 2\pi i b^{\mp 1}(\Z e_1+\Z e_2)$. Assuming the value of $b^2$ to be non-rational, this restricts $s$ to a one-dimensional space. For example, the pair
\bea
s=\kappa \om_1 + \pi i (\ell b+m b^{-1})e_2 \scs s'=\kappa \om_1 +\pi i (\ell b-mb^{-1}) e_2 \scs (\kappa,\ell,m)\in \R\times \N^2
\eea
is such that $s-s'=2\pi i mb^{-1}e_2$ and $s-r(s')=2\pi i \ell b e_2$. Therefore,   
\bea
U_{\kappa|\ell,m}(\al)&\equiv &
U_{\kappa \om_1 + \pi i (\ell b+m b^{-1})e_2}(\al) -U_{\kappa \om_1 +\pi i (\ell b-mb^{-1}) e_2}(\al) 
\label{udiff}
\eea
is a solution of eq. (\ref{lau}), and an explicit calculation yields
\bea
\boxed{ U_{\kappa|\ell,m}(\al) = \frac{4}{A(\al)} \sum_{e>0} e^{\kappa(h_e,\al-Q)} \sin 2\pi \ell b(e,\al-Q) \sin 2\pi m b^{-1}(e,\al-Q) } \ ,
\label{uess}
\eea
where to a positive root $e$ we associate the weight $h_e\in H_{\om_1}$ such that $(e,h_e)=0$. In this expression the sine factors compensate some of the $\al-Q\rar 0$ divergences of the prefactor $A(\al)$ (\ref{aal}). We hold that $U_{\kappa|\ell,m}(\al)$ with $\kappa \in \R$ and $\ell,m\in \N$ define the family of simply degenerate D-branes. The boundary parameters of such D-branes are 
\bea
\lambda_{1,+}=e^{-\frac23 b\kappa} +2e^{\frac13 b\kappa}(-1)^m \cos\pi\ell b^2 \scs \lambda_{1,-}=e^{-\frac23 b^{-1}\kappa} +2e^{\frac13 b^{-1}\kappa}(-1)^\ell \cos\pi m b^{-2}\ ,
\label{lpm}
\eea
and $\lambda_{2,\pm}$ are obtained by $\kappa \rar -\kappa$. 

We expect that in the more general $s\ell_n$ case this construction generalizes to a hierarchy of partly degenerate D-branes. The difference of two terms in eq. (\ref{udiff}) should be interpreted as a sum over the $\Z_2$ subgroup of the Weyl group which leaves $\om_1$ invariant, weighted by the signatures of the elements of that subgroup. In $s\ell_n$, there is a hierarchy of subgroups of the Weyl group which leave certain hyperplanes invariant, and summing over such subgroups should yield the partly degenerate D-branes. The case of fully degenerate D-branes corresponds to a sum over the full Weyl group, which we now study in the $s\ell_3$ case. 

Given two integral dominant weights $\Omega,\Omega'\in \N\om_1+\N\om_2$ we consider the combination
\bea
U_{\Omega|\Omega'}(\al)\equiv \sum_{w\in W} \e(w)\ U_{2\pi i(b(\Omega+\rho)+b^{-1}w(\Omega'+\rho))}(\al)\ .
\label{uoo}
\eea
This can be rewritten as 
\bea
\boxed{U_{\Omega|\Omega'}(\al)=\frac{1}{A(\al)} \sum_{w\in W} \e(w)\ e^{2\pi i b(w(\Omega+\rho),\al-Q)} \sum_{w'\in W} \e(w')\ e^{2\pi i b^{-1}(w'(\Omega'+\rho),\al-Q)}}\ .
\label{uww}
\eea
This can be shown to have a finite limit as $\al-Q\rar 0$, by first using the Weyl formula (\ref{cop}) in order to reduce the problem to the case of $U_{0|0}(\al)$, 
\bea
\frac{U_{\Omega|\Omega'}(\al)}{U_{0|0}(\al)} = \chi_{\Omega}(2\pi i b(\al-Q))\ \chi_{\Omega'}(2\pi ib^{-1}(\al-Q))\ ,
\label{uucc}
\eea
and then the formula (\ref{swpe}) in order to prove the regularity of $U_{0|0}(\al)$. 
We interpret $U_{\Omega|\Omega'}(\al)$ as defining localized (zero-dimensional) D-branes, which we will call discrete D-branes or fully degenerate D-branes. But first we should check that the six values of $s$ which are involved in the sum (\ref{uoo}) do have the same boundary parameters $\lambda_{i,\pm}$ (\ref{lch}). This is actually true, as a consequence of the fact that the weights $h\in H_{\om_1}$ differ from one another by elements of $\Z e_1+\Z e_2$. And we find
\bea
\lambda_{1,+}=e^{-2\pi i(\om_1,\Omega')}\chi_{\om_1}(-2\pi i b^2\Omega) \scs \lambda_{1,-}=e^{-2\pi i(\om_1,\Omega)}\chi_{\om_1}(-2\pi i b^{-2}\Omega')\ .
\label{lloo}
\eea
In the case of discrete D-branes, these boundary cosmological constants are expected to be directly related to the values $U_s(-b^{\pm 1}\om_1)$ of the one-point structure constant, as was argued in the case of Liouville theory in \cite{zz01}. For example, $\lambda_{1,+}$ originally appeared as the bulk-boundary structure constant $R_0$ in the bulk-boundary OPE of $V_{-b\om_1}$ (\ref{vrb}). If the one-point structure constant was normalized so that $U_{\Omega|\Omega'}(0)=1$, then $\lambda_{1,+}$ would coincide with $U_{\Omega|\Omega'}(-b^{\pm 1}\om_1)$. Therefore we expect
\bea
\lambda_{i,\pm}=\frac{U_{\Omega|\Omega'}(-b^{\pm 1}\om_i)}{ U_{\Omega|\Omega'}(0)}\ .
\label{luu}
\eea
And indeed the expressions (\ref{uww}) and (\ref{lloo}) obey such relations, as can be shown with the help of the Weyl formula (\ref{cop}). (For this to be completely true we would have to reinstate the simple factor $\frac{\G(-2-3b^2)}{\G(-b^2)} \left[-\frac{\pi \mu}{\g(-b^2)}\right]$ in $\lambda_{i,\pm}$; such a factor was present in eq. (\ref{fc}) but we neglected it in what followed.)
It is amusing to note that we did already apply the Weyl formula to $U_{\Omega|\Omega'}(\al)$ before, but in a different, ``dual'' way, in order to prove eq. (\ref{uucc}). 

Our expression for the bulk one-point function can be used to show that the equations of motion derived from the Lagrangian $L_n$ (\ref{ln}) of Toda theory are obeyed in the presence of discrete D-branes. If we identify $e^{(\al,\phi)}$ with the operator $V_\al$, and therefore $\phi_i$ with $\left.\pp{\al_i} \right|_{\al=0} V_\al$, then 
the quantum version of the equations of motion is 
\bea
\left.\pp{\al_i} \right|_{\al=0} \p\bp \la V_\al(z) \ra = \pi b \mu \la V_{be_i}(z) \ra\ .
\eea
Using the form $\la V_\al(z) \ra = \frac{U(\al)}{|z-\bz|^{2\Delta_\al}}$ (\ref{vu}) of the one-point function, this reduces to the following identity for the structure constant $U(\al)$:
\bea
\frac{U(be_i)}{U(0)}=\frac{2(b+b^{-1})}{\pi b \mu} \ .
\eea
It can be checked that this identity is obeyed by the one-point structure constant $U_{\Omega|\Omega'}(\al)$ (\ref{uww}) of a discrete D-brane.

\subsection{Modular bootstrap analysis \label{ssmo}}

The factorization constraint (\ref{vvcug})-(\ref{vvbbf}) whose solutions were our one-point structure constants is only one of the many equations of the conformal bootstrap formalism. Another one of these equations is relatively tractable: the modular bootstrap constraint, which relates two different decompositions of the annulus partition function
\bea
Z_{s_1;s_2}(\tau) &=& \frac{1}{6}\int d^2p\ \ U_{s_1}(Q+ip)U_{s_2}(Q-ip)\ \xi_{Q+ip}(\tau)\ ,
\label{anc}
\\
&=& {\rm Tr}_{H_{s_1;s_2}}\ e^{-\frac{2i\pi}{\tau}\left(L_0-\frac{c}{24}\right)}\ .
\label{ano}
\eea
Let us explain these formulas. We consider an annulus, the simplest Riemann surface with two boundaries. The two boundaries are characterized by their boundary parameters $s_1,s_2$, which may each correspond to any type of D-brane: continuous, simply degenerate, or discrete; and the geometry of the annulus is characterized by the modular parameter $\tau$. The annulus partition function (or zero-point correlation function) $Z_{s_1;s_2}(\tau)$ first has a ``bulk channel'' decomposition, which describes the exchange of bulk operators between the two boundaries. The resulting formula (\ref{anc}) for $Z_{s_1;s_2}(\tau)$ therefore involves a sum over the bulk spectrum. This sum decomposes into an integral over the physical values (\ref{spec}) of the momenta $\al=Q+ip$ which characterize the highest-weight representations of the algebra $W_3$ (with the factor $\frac16$ due to the Weyl symmetry) , and sums over the descendent states in each representation, which are encoded in the characters $\xi_\al(\tau)$. The one-point structure constants $U_{s_1}(Q+ip)$ and $U_{s_2}(Q-ip)$ involve ingoing and outgoing momenta respectively, and no normalization factors appear due to the normalization assumption of \cite{fl07c}, $\la V_{Q+ip_1}(z_1) V_{Q-ip_2}(z_2) \ra = \frac{\delta(p_1-p_2)}{|z_{12}|^{4\Delta_{Q+ip_1}}}$ (where we write only one of the six terms of a sum over the Weyl group).

The annulus partition function also has a ``boundary channel'' decomposition, which describes a one-loop partition function of open strings. The resulting formula (\ref{ano})  for $Z_{s_1;s_2}(\tau)$ is the trace over the boundary spectrum $H_{s_1;s_2}$ of the propagator $e^{-\frac{2i\pi}{\tau}\left(L_0-\frac{c}{24}\right)}$. This propagator is the operator which appears in the definition (\ref{xtr}) of the characters $\xi_\al(\tau)$; in the boundary channel it appears with the dual value $-\frac{1}{\tau}$ of the modular parameter. Although we do not know the boundary spectrum $H_{s_1;s_2}$, the modular bootstrap method will produce tests of the one-point structure constants $U_s(\al)$. This is because for some choices of $s_1,s_2$ the spectrum is discrete, and the requirement that each representation should appear with a positive integer multiplicity is a nontrivial constraint. 

Let us compute $Z_{s_1;s_2}(\tau)$ by using the bulk channel decomposition (\ref{anc}) with the bulk structure constants $U_s(\al)$ which we found in the previous two Subsections. We start with the case when both D-branes are discrete, and use eq. (\ref{uucc}):
\begin{multline}
Z_{\Omega_1|\Omega_1';\Omega_2|\Omega_2'}(\tau)=\frac16\int d^2p\ \xi_{Q+ip}(\tau)\ U_{0|0}(Q+ip)U_{0|0}(Q-ip) 
\\ \times 
\chi_{\Omega_1^*}(2\pi b p) \chi_{\Omega_2}(2\pi bp) \chi_{\Omega_1'^*}(2\pi b^{-1}p)\chi_{\Omega_2'}(2\pi b^{-1}p)\ ,
\end{multline}
where we used the property $\chi_{\Omega}(-p)=\chi_{\Omega^*}(p)$. Now let us decompose the products of characters using eq. (\ref{rrr}), while computing the product $U_{0|0}(Q+ip)U_{0|0}(Q-ip)$ using eq. (\ref{swpe}),
\begin{multline}
Z_{\Omega_1|\Omega_1';\Omega_2|\Omega_2'}(\tau)=\frac{(2\pi b^{-1})^6}{6}\sum_{\Omega}m_{\Omega_1^*,\Omega_2}^{\Omega} \sum_{\Omega'} m_{\Omega_1'^*,\Omega_2'}^{\Omega'}\int d^2p\ \xi_{Q+ip}(\tau)\ 
\\ \times 
\chi_{\Omega}(2\pi bp)\chi_{\Omega'}(2\pi b^{-1}p) \prod_{e>0}\prod_\pm 
(e^{\frac12(e,2\pi b^{\pm 1}p)}-e^{-\frac12(e,2\pi b^{\pm 1}p)})\ .
\end{multline}
The value of the (Gaussian) integral is given by eq. (\ref{xoo}),
\bea
Z_{\Omega_1|\Omega_1';\Omega_2|\Omega_2'}(\tau)=\frac{(2\pi b^{-1})^6}{\sqrt{3}}\sum_{\Omega,\Omega'}m_{\Omega_1^*,\Omega_2}^{\Omega}  m_{\Omega_1'^*,\Omega_2'}^{\Omega'}\ \xi_{-b\Omega-b^{-1}\Omega'}(-\tfrac{1}{\tau})\ .
\eea
This is a sum of characters with positive integer coefficients $m_{\Omega_1^*,\Omega_2}^{\Omega}m_{\Omega_1'^*,\Omega_2'}^{\Omega'}$, up to a factor which could be absorbed in a renormalization of the one-point structure constant $U_{\Omega|\Omega'}(\al)$. The characters are those of fully degenerate representations. As we pointed out in Subsection \ref{ssrep}, characters $\chi_\al(\tau)$ do not fully characterize representations of the $W_3$ algebra, nevertheless we conjecture that the boundary spectrum is $H_{\Omega_1|\Omega_1';\Omega_2|\Omega_2'}=\oplus_{\Omega,\Omega'}m_{\Omega_1^*,\Omega_2}^{\Omega}  m_{\Omega_1'^*,\Omega_2'}^{\Omega'}\ R_{-b\Omega-b^{-1}\Omega'}$. If we associate the representation $R_{-b\Omega-b^{-1}\Omega'}$ to the discrete D-brane of parameter $s=\Omega|\Omega'$, 
fusing the representations associated to the two involved D-branes (after conjugating one of them) produces the boundary spectrum, which agrees with Cardy's ideas. 

In particular, the $0|0$ D-brane corresponds to the identity representation, and for any D-brane of parameter $s$, the spectrum $H_{0|0;s}$ should be the single representation which is associated to that D-brane. 
In the case of a continuous D-brane,
\bea
Z_{0|0;s}(\tau) =\frac{(2\pi b^{-1})^6}{6} \int d^2p\ \xi_{Q+ip}(\tau)\ \sum_{w\in W} e^{i(w(s),p)} = \frac{(2\pi b^{-1})^6}{\sqrt{3}}\xi_{Q-\frac{s}{2\pi i}}(-\tfrac{1}{\tau})\ ,
\eea
so that the continuous representation of momentum $Q-\frac{s}{2\pi i}$ is associated to the continuous D-brane of parameter $s$. This immediately generalizes to simply degenerate D-branes, using the formula (\ref{udiff}) for their one-point structure constants. Another generalization is to replace the identity D-brane with an arbitrary discrete D-brane,
\bea
Z_{\Omega|\Omega';s}(\tau)&=&\frac{(2\pi b^{-1})^6}{6} \int d^2p\ \xi_{Q+ip}(\tau)\ \chi_{\Omega^*}(2\pi b p) \chi_{\Omega'^*}(2\pi b^{-1}p)\sum_{w\in W} e^{i(w(s),p)} 
\\
&=& \frac{(2\pi b^{-1})^6}{\sqrt{3}}\sum_{h\in H_\Omega}\sum_{h'\in H_{\Omega'}} \xi_{Q-\frac{s}{2\pi i}-bh-b^{-1}h'}(-\tfrac{1}{\tau})\ ,
\eea
which lends support to the conjecture that the boundary spectrum is obtained by fusing the representations which correspond to the two D-branes. (Notice that a weight $h$ may appear several times in $H_\Omega$, as happened in the definition of the character $\chi_\Omega(p)$ (\ref{chio}).)

\subsection{Continuous boundary spectra \label{ssspec}}

Let us investigate the boundary spectrum of $s\ell_n$ Toda theory in the presence of two continuous D-branes using the modular bootstrap approach. Up to numerical factors, the annulus partition function is formally written as
\bea
Z_{s_1;s_2}(\tau) = \int d^{n-1}p\  \frac{\sum_{w_1,w_2\in W} e^{i(w_1(s_1)-w_2(s_2),p)}}{\prod_\pm \prod_{e>0} \sinh \pi b^{\pm 1}(e,p) }\ \xi_{Q+ip}(\tau)\  .
\label{zsst}
\eea
This expression suffers from infrared divergences near $p=0$. In the case of Liouville theory ($n=2$), the divergence is of the type $\int \frac{dp}{p^2}$ and therefore linear in a long distance cutoff $L$. This is attributed to the geometry of the continuous D-branes, which are supposed to extend up to infinity in the Liouville space of coordinate $\phi$. One may therefore naively expect that for general $n$ the divergence should be $L^{n-1}$. However, it is actually $L^{(n-1)^2}$, so there is an extra divergence $L^{(n-1)(n-2)}$ beyond what is expected on geometrical grounds. 

We observe that this extra divergence is governed by the number $2d_n=(n-1)(n-2)$ of parameters which are necessary to account for the infinite fusion multiplicities of continuous representations, where the factor $2$ in $2d_n$ is meant to take into account the antiholomorphic multiplicities. (See Subsection \ref{ssmul}.)  Combining this observation with Cardy's ideas suggests a heuristic explanation of the divergence. Indeed, if the boundary spectrum is obtained by fusing the representations associated to the two D-branes, and fusion multplicity is infinite, then the boundary spectrum is a sum of representations with infinite multiplicities.
This must then lead to divergences in the annulus partition function, in addition to the ordinary ``geometrical'' divergences.
Therefore, we conjecture that the boundary spectrum in the presence of two continuous D-branes is the sum of all continuous representations of the $W_n$ algebra, each one appearing with an infinite multiplicity.

It is not obvious to us how these considerations generalize to annulus partition functions involving arbitrary D-branes. For example,
in the case of $s\ell_3$ Toda theory with one continuous and one simply degenerate D-brane, the annulus partition function diverges as $L^2$. The fusion multiplicity is finite in this case, and we conjecture that the spectrum is the sum of all continuous representations, each one appearing a finite number of times.

\zeq\section{Light asymptotic limits of some correlation functions \label{secla}}

The main purpose of this Section is to establish a link between the classical analysis of Section \ref{sted}, and the conformal bootstrap analysis of Section \ref{secbo}, in the case of $s\ell_3$ Toda theory with the boundary condition $W+\bar{W}=0$. We will use the classical solutions of the Toda equations for predicting the bulk one-point function in a certain limit, and we will find that the classical predictions agree with the bootstrap results up to unimportant details. In addition we will also predict the light asymptotic limits of the boundary two-point function in the case $W+\bar{W}=0$, and of bulk one-point and boundary two-point functions in the case $W-\bar{W}=0$, for which we did not perform the conformal bootstrap analysis. 
For pedagogical purposes, we will begin with the computation of the analogous correlation functions in the much simpler case of Liouville theory.

The light semi-classical asymptotic limit, or light asymptotic limit, of a correlation function $\la \prod_i V_{\al_i}(z_i)\ra_s$ in the presence of a boundary with parameter $s$, where $V_{\al_i}(z_i)$ may be a bulk or a boundary operator, is defined by
\bea
b\rar 0 \scs \eta_i\equiv b^{-1}\al_i\ {\rm and}\ \sigma\equiv bs\ {\rm fixed}\ .
\label{bes}
\eea
If the correlation function is formally represented as a functional integral over the Toda field $\phi$, with the weight $e^{-S[\phi]}$ where $S[\phi]$ is the action, then the light asymptotic limit reduces that functional integral to a finite-dimensional integral over field configurations such that $W^{(s)}=\bar{W}^{(s)}=0$ for all spins $s$ \cite{fl07c}. In $s\ell_n$ Toda theory, such field configurations are those where $X_i=e^{-\phi_i}$ are polynomials of degree $n-1$ as functions of the coordinates $z,\bz$. These polynomials must obey further constraints like the boundary conditions and the reality of $\phi_i$. The functional integral reasoning therefore predicts that the quantity
\bea
\la \prod_i e^{(\eta_i,\phi(z_i))} \ra_{\sigma}^{light} \equiv \int_{M_\sigma}d\phi_i\ \prod e^{(\eta_i,\phi(z_i))}\ ,
\label{els}
\eea
where ${M_\sigma}$ is a finite-dimensional space of field configurations, should be related to the limit (\ref{bes}) of $\la \prod_i V_{\al_i}(z_i)\ra_s$,
\bea
\underset{b\rar0}{\lim} \frac{\la \prod_i V_{b\eta_i}(z_i)\ra_{b^{-1}\sigma}}{\la V_0(0) \ra_{b^{-1}\sigma}}=\frac{\la \prod_i e^{(\eta_i,\phi(z_i))} \ra_{\sigma}^{light}}{\la 1 \ra_{\sigma}^{light}}
\ .
\label{vei}
\eea
Here we normalize the correlation functions by dividing them by the partition function. This eliminates the dependences on the overall undetermined factor in the integration measure on $M_\sigma$, and on the value $S[\phi]$ of the action for the polynomial field configurations, which is difficult to compute. As a result, nothing in eq. (\ref{vei}) depends on the action, and we conjecture that that equation holds 
whether a boundary action exists or not. (See Appendix \ref{secac} for a discussion of that point.)

\subsection{Case of Liouville theory}

Let us consider the solutions (\ref{xsol}) of the Liouville equation which obey the ``light asymptotic condition'' 
$\p^2X=\bp^2X=0$, where $X=e^{-\phi}$. Such solutions are built from two functions $b_1,b_2$ which are polynomials of degrees at most one. Given the freedom to choose the matrix $N$, we can fix these functions without loss of generality, and we choose $b_1(z)=1,\ b_2(z)=z$. So we have $X(z)=(1,\bz)N\bsm 1\\z\esm$ where $N$ is a positive Hermitian matrix of determinant one. 
According to equation (\ref{els}), the one-point function of a bulk operator in the presence of a boundary with parameter $\lambda_L$ (\ref{trnp}) is of the type
\bea
\la e^{\eta \phi(z)} \ra_{\lambda_L}^{light} =  \int dN\ \delta\left(\lambda_L-\tfrac{1}{2i}\Tr(NP)\right)\ X(z)^{-\eta}\ ,
\label{vesl}
\eea
where $dN$ is an integration measure, and we expect the boundary parameters $\lambda_L$ and $\sigma$ to be related. 

Assuming that the integration measure $dN$ is invariant under the 
$\SLR$ symmetry $N\rar \Lambda^T N\Lambda$ (\ref{linv}), let us show that this symmetry determines
the $z$-dependence of the one-point function. The subgroup of $\SLR$ which survives our fixing of $b_1,b_2$ is the set of matrices of the type $\Lambda(x,y)\equiv\bsm \sqrt{y} & 0 \\ -\frac{x}{\sqrt{y}} & \frac{1}{\sqrt{y}} \esm$, which are such that
$\Lambda(x,y) \bsm 1 \\ x+iy \esm =\sqrt{y} \bsm 1 \\ i \esm  $. By using this residual subgroup in eq. (\ref{vesl}) we obtain $\la e^{\eta \phi(z)} \ra_{\lambda_L}^{light} = (\Im z)^{-\eta}\la e^{\eta \phi(i)} \ra_{\lambda_L}^{light} $, which agrees with what we would expect from conformal symmetry.

There is a simple method for computing the integral in eq. (\ref{vesl}), which unfortunately does not easily generalize to $s\ell_{n>2}$ Toda theory. It uses the parametrization $N=\bsm Y_0-Y_1 & Y_2+iY_3 \\ Y_2-iY_3 & Y_0+Y_1\esm$, where the constraint $\det N=1$ still has to be imposed. We have
\bea
\la e^{\eta \phi(i)} \ra_{\lambda_L}^{light}  =  \int \left(\prod_{i=0}^3 dY_i\right)\ \delta(\lambda_L+Y_3)\ \delta(Y_0^2-Y_1^2-Y_2^2-Y_3^2-1)\ (2Y_0-2Y_3)^{-\eta} \ .
\eea
Integrating over $Y_3$ then $Y_1$ and $Y_2$, we obtain
$
\la e^{\eta \phi(i)} \ra_{\lambda_L}^{light}  =  \frac{\pi}{2^\eta} \int_{\sqrt{\lambda_L^2+1}}^\infty dY_0 (Y_0+\lambda_L)^{-\eta}
$
from which we deduce
\bea
\la e^{\eta \phi(z)} \ra_{\lambda_L}^{light}=(\Im z)^{-\eta}  \frac{\pi}{2^\eta} \frac{\left(\lambda_L+\sqrt{\lambda_L^2+1}\right)^{1-\eta}}{\eta-1}\ .
\label{vell}
\eea
Let us now present another calculation of the integral (\ref{vesl}), which can more easily be generalized to the case of $s\ell_n$ Toda theory with $n\geq 3$. We adopt the parametrization $N=\bar{M}^TM=\bsm \rho^{-2}+|a|^2 & \rho \bar{a} \\ \rho a & \rho^2 \esm$ where $M= \bsm \rho^{-1} & 0 \\ a & \rho\esm$ is a function of a real parameter $\rho$ and a complex parameter $a$. The integral (\ref{vesl}) becomes
\bea
\la e^{\eta \phi(z)} \ra_{\lambda_L}^{light} =  \int \rho d\rho\ d^2a\ \delta(\lambda_L+\rho\Im a)\ \left(\rho^{-2} +|a+\rho z|^2\right)^{-\eta}\ .
\eea
It is possible, but tedious, to compute this integral directly. Instead, let us introduce the notation
\bea 
\lambda_L=i\cosh  \sigma\ .
\label{llss}
\eea
For the moment this is a rather awkward notation, as we have to assume that $ie^\sigma$ is real (and we further take it to be positive). In addition we perform the change of variables $a\rar a-\rho z$, and we obtain
\bea
\la e^{\eta \phi(z)} \ra_{\sigma}^{light}=  \int \rho d\rho\ d^2a\ \delta\left(\tfrac{i}{2} e^{\sigma}+\tfrac{i}{2} e^{-\sigma}+\rho\Im a-\rho^2\Im z\right)\ \left(\rho^{-2} +|a|^2\right)^{-\eta}\ .
\eea
In the limit
\bea
\bla
ie^{\sigma} \rar \xi ie^{\sigma} \\ z\rar \xi z \ela \scs \xi\rar \infty\ ,
\label{limxi}
\eea
the integral greatly simplifies, 
\bea
\la e^{\eta \phi(z)} \ra_{\sigma}^{light} \sim \frac{1}{\xi\Im z}\   \int d^2a\ \left(\tfrac{2\Im z}{ie^\sigma} +|a|^2\right)^{-\eta} = (\xi \Im z)^{-\eta}   \frac{\pi}{2^\eta} \frac{\left(i\xi e^{\sigma}\right)^{\eta-1}}{\eta-1}\ . 
\label{eeps}
\eea
This agrees with the result (\ref{vell}), in spite of the limit (\ref{limxi}) which we have taken. This is because the result (\ref{vell}) has a very simple behaviour under the rescalings involved in the definition of the limit. The behaviour under the rescaling of $z$ is of course a consequence of conformal symmetry, but we have no a priori reason for the behaviour under the rescaling of $ie^{ \sigma}$ to be simple. We will assume that a similar behaviour persists in $s\ell_n$ Toda theory with arbitrary $n$, and this will allow us to take limits analogous to (\ref{limxi}) before performing integrals which would otherwise seem intractable.

Now let us investigate whether the relation (\ref{vei}) between the classical and quantum calculations of the one-point function holds. According to eq. (\ref{eeps}), the normalized classical one-point function is  
\bea
\frac{\la e^{\eta \phi(z)} \ra_{\sigma}^{light}}{\la 1 \ra_{\sigma}^{light}} &=& (2\Im z)^{-\eta} \frac{(ie^{\sigma})^{\eta}}{1-\eta} \ .
\eea
The behaviour of the exact Liouville one-point function (\ref{opt}) in the light asymptotic limit is 
\bea
\la V_{b\eta}(z)\ra_{b^{-1}\sigma} &\underset{b\rar 0}{\sim}& (2\Im z)^{-\eta} (\pi b^{-2}\mu)^{-\frac{\eta}{2}}\G(\eta-b^{-2})\ e^{|\sigma| b^{-2}}\ \frac{e^{|\sigma| (1-\eta)}}{\eta-1}\ ,
\label{vllim}
\\
\frac{\la V_{b\eta}(z)\ra_{b^{-1}\sigma}}{\la 1 \ra_{b^{-1}\sigma}} &\underset{b\rar 0}{\sim}& (2\Im z)^{-\eta}(-1)^\eta (\pi b^{2}\mu)^{-\frac{\eta}{2}} \frac{e^{-\eta|\sigma|}}{1-\eta}\ ,
\eea
where we assume $\sigma$ to be real, and we used $\frac{\G(\eta-b^{-2})}{\G(-b^{-2})}\underset{b\rar 0}{\sim} (-b^{-2})^\eta$. The classical and quantum results agree, provided 
the cosmological constant takes the value $\mu=-\frac{1}{\pi b^2}$ (\ref{muval}) which was assumed in the classical analysis, and the classical boundary parameter $\sigma$ defined in eq. (\ref{llss}) is identified with the quantum boundary parameter $\sigma$ which appears in eq. (\ref{vllim}). 
However, $ie^\sigma$ is assumed to be real in the classical analysis, whereas $\sigma$ is assumed to be real in the calculation of the limit of the exact one-point function. Thus, the comparison between the two methods must involve an analytic continuation of the results. This problem ultimately comes from the fact that the bulk cosmological constant $\mu$ is assumed to be negative in the classical analysis, and positive in the conformal bootstrap analysis. And the known relation \cite{fzz00} between the boundary cosmological constant $\lambda_L$ and $\sigma$ is in our notations $\left(\frac{\lambda_L}{\pi b^2}\right)^2 =\frac{\mu}{\sin \pi b^2} \cosh^2 \sigma$, which agrees with eq. (\ref{llss}) in the $b\rar 0$ limit if $\mu=-\frac{1}{\pi b^2}$. 
This confirms our earlier identification of $\lambda_L$ as the boundary cosmological constant, see eq. (\ref{ltb}).

Finally, let us compute the light asymptotic limit of the boundary two-point function:
\bea
\la e^{\eta_1\phi(z_1)}e^{\eta_2\phi(z_2)} \ra_{\lambda_L}^{light} =   \int \rho d\rho\ d^2a \ \delta(\lambda_L+\rho\Im a)\ \prod_{i=1}^2\left(\rho^{-2} +|a+\rho x_i|^2\right)^{-\eta_i}\ .
\eea
Let us use conformal invariance, and fix $x_1=0,x_2=\infty$. This makes the computation elementary, and the result is
\bea
\la e^{\eta_1\phi(0)}e^{\eta_2\phi(\infty)} \ra_{\lambda_L}^{light} =   \delta(\eta_1-\eta_2)\ \frac{\G(\eta_1-\frac12)}{\G(\eta_1)} (\lambda_L^2+1)^{\frac12-\eta_1}\ .
\eea
This agrees with the expectations from the conformal bootstrap analysis \cite{fzz00}, provided our relation (\ref{llss}) between the classical and quantum boundary parameters is assumed.

\subsection{Case of $s\ell_3$ Toda theory with $W-\bar{W}=0$}

Let us consider the solutions (\ref{xmsol}) of the $s\ell_3$ Toda equations which obey the ``light asymptotic condition'' $\p^3X_i=\bp^3X_i=0$. 
Such solutions are built from three functions $b_1,b_2,b_3$ which are polynomials of degrees at most two. Given the freedom to choose the matrix $N$, we can fix these functions without loss of generality, and we choose $B(z)\equiv (b_1(z),b_2(z),b_3(z))=(\frac12 z^2,z,1)$. The corresponding Wronskians are $(w_1(z),w_2(z),w_3(z))=(1,-z,\frac12 z^2)$ and they obey $w_i=P_{ij}b_j$ with $P\equiv\bsm 0 & 0 & 1 \\ 0 & -1 & 0 \\ 1 & 0 & 0\esm$. This leads to
\bea
X_1=B(\bz) N B(z)^T \scs X_2= B(\bz) P N^{-1T} P B(z)^T\ .
\eea
According to equation (\ref{els}), the one-point function of a bulk operator $e^{(\eta,\phi(z))}=e^{\eta_1\phi_1(z)+\eta_2\phi_2(z)}$ in the presence of a boundary with parameter $\lambda_0$ (\ref{ldn}) is of the type
\bea
\la e^{(\eta,\phi(z))} \ra_{\lambda_0}^{light} =  \int dN\ \delta\left(\lambda_0-\det \tfrac12(N+N^T)\right)\ X_1^{-\eta_1}X_2^{-\eta_2}\ .
\label{vml}
\eea

We may use the $SL(3,\R)$ symmetry $N\rar \Lambda N\Lambda^T$ (\ref{tminv}) in order to determine the dependence of the one-point function on $z=x+iy$. Introducing the family of matrices
$\Lambda(x,y) \equiv \bsm y^{-1} & -xy^{-1} & \frac12 x^2 y^{-1} \\ 0 & 1 & -x  \\ 0 & 0 & y \esm$, we have $\Lambda(x,y)^TP\Lambda(x,y)=P$ and $\Lambda(x,y)B(x+iy)^T=yB(i)^T$. Together with the assumption that the measure $dN$ in eq. (\ref{vml}) is invariant under the symmetry, this implies $\la e^{(\eta,\phi(z))} \ra_{\lambda_0}^{light}=(\Im z)^{-2\eta_1-2\eta_2} \la e^{(\eta,\phi(i))} \ra_{\lambda_0}^{light}$. The power of $\Im z$ thus obtained is $-2(\rho,\eta)=\underset{b\rar 0}{\lim}( -2\Delta_{b\eta})$ where $\Delta_\al$ given by eq. (\ref{da}), as expected from conformal symmetry. Now it turns out that the $SL(3,\R)$ symmetry can yield further information on the one-point function. This is because after assuming $z=i$ there still is a residual subgroup of matrices $\Lambda(d,e,f)\equiv \bsm f & -2d & 2f-2e \\ d & e & -2d \\ \frac12 f-\frac12 e & d & f \esm$ where $d,e,f$ are real parameters constrained by $\det\Lambda(d,e,f)=(e^2+4d^2)(2f-e)=1$, and such matrices obey the relations $\Lambda(d,e,f)^{-1T} B(i)^T=\frac{1}{e-2id}B(i)^T$ and $\Lambda(d,e,f)PB(i)^T=(e+2id)PB(i)^T$. Thus, under transformations $N\rar \Lambda^{-1}N\Lambda^{-1T}$, we have $X_1(i)\rar \frac{1}{e^2+4d^2} X_1(i)$ and $X_2(i)\rar (e^2+4d^2)X_2(i)$. This shows that $\la e^{(\eta,\phi(i))} \ra_{\lambda_0}^{light}$ must vanish unless $\eta_1=\eta_2$. The consequences of the $SL(3,\R)$ symmetry may be summarized as
\bea
\la e^{(\eta,\phi(z))} \ra_{\lambda_0}^{light}=\delta(\eta_1-\eta_2)\ (\Im z)^{-4\eta_1} \la e^{\eta_1(\rho,\phi(z))} \ra_{\lambda_0}^{light}\ .
\eea

Let us now introduce a parametrization $N=\bar{M}^TM$ in terms of a triangular matrix $M=\bsm \rho & a & b \\ 0 & \nu & c \\ 0 & 0 & \tau \esm$ which depends on three real parameters $\rho,\nu,\tau$ such that $\rho\nu\tau=1$, and three complex parameters $a,b,c$. Then
$N=\bsm \rho^2 & \rho a & \rho b \\ \rho \bar{a} & \nu^2 + |a|^2 & \bar{a}b+\nu c \\ 
\rho \bar{b} & a\bar{b} + \nu\bar{c} & |b|^2+|c|^2+\tau^2 \esm$ and 
\bea
dN&=& \tfrac{1}{\pi^3}\rho^3 d\rho\ \nu d\nu\ d^2a\ d^2b\ d^2c\ ,
\label{dnt}
\\
X_1&=&\tau^2 + |c+\nu z|^2 +|b+az+\tfrac12 \rho z^2|^2 \ ,
\label{xone}
\\
X_2 &=& \rho^{-2} + |\tau a +\nu^{-1} z|^2 +|ac-\nu b+ \rho c z +\tfrac12 \tau^{-1}z^2|^2\ .
\label{xtwo}
\eea
In terms of such variables, the boundary parameter is
\bea
\lambda_0 = \det\tfrac12(N+N^T)= \rho^2\left[(\nu \Im b - \Re c\Im a)^2+(\nu^2+\Im a^2 )(\tau^2+\Im c^2)\right]\ .
\eea
We consider the one-point function (\ref{vml}) and perform the simultaneous shifts 
\bea
a\rar a-\rho z\scs b\rar b-az+\tfrac12\rho z^2 \scs c\rar c-\nu z\ ,
\label{shift}
\eea
thereby obtaining
\begin{multline}
\la e^{(\eta,\phi(z))} \ra_{\lambda_0}^{light}= \int \frac{dN}{\rho^2}\ (\tau^2+|b|^2+|c|^2)^{-\eta_1}(\rho^{-2}+\tau^2|a|^2+|ac-\nu b|^2)^{-\eta_2} 
\\ \delta\left(\frac{\lambda_0}{\rho^2}-[\nu\Im b-\Re c\Im a+\Im z(\rho \Re c-\nu\Re a)]^2-[\nu^2+(\Im a-\rho \Im z)^2][\tau^2+(\Im c-\nu \Im z)^2] \right) \ .
\end{multline}
In the limit
\bea
\bla
\lambda_0 \rar \xi^4\lambda_0 \\
z\rar \xi z\ela \scs \xi\rar \infty\ ,
\label{lzlim}
\eea
the integral simplifies and reduces to
\bea
\la e^{(\eta,\phi(z))} \ra_{\lambda_0}^{light} &\sim& \frac{1}{\xi^4}(\Im z)^{-2\eta_1-2\eta_2} \lambda_0^{\frac{\eta_1+\eta_2-2}{2}} \int \frac{d\nu}{\nu} \nu^{\eta_1-\eta_2} \times I_{\eta_1,\eta_2}\ ,
\\
I_{\eta_1,\eta_2}&\equiv &
\tfrac{1}{\pi^3}\int d^2a\ d^2b\ d^2c\ (1+|b|^2+|c|^2)^{-\eta_1} (1+|a|^2+|ac-b|^2)^{-\eta_2}\ .
\label{iee}
\eea
The integral $I_{\eta_1,\eta_2}$ can be computed for arbitrary values of $\eta_1,\eta_2$ by making repeated use of the formula $\frac{1}{\pi}\int d^2z\ (|z|^2+A)^{-\eta} = \frac{A^{1-\eta}}{\eta-1}$, with the result
\bea
I_{\eta_1,\eta_2} =  \frac{1}{(\eta_1-1)(\eta_2-1)(\eta_1+\eta_2-2)}=\prod_{e>0}\frac{1}{(e,\eta-\rho)}\ .
\label{ieer}
\eea
Using this result in the particular case $\eta_1=\eta_2$, we obtain the expression for the one-point function in the limit (\ref{lzlim}),
\bea
\xi^4 \la e^{(\eta,\phi(z))} \ra_{\lambda_0}^{light} &\sim& \delta(\eta_1-\eta_2)\ (\Im z)^{-4\eta_1} \frac{\lambda_0^{\eta_1-1}}{(\eta_1-1)^3}\ .
\label{eet}
\eea
By analogy with the case of Liouville theory, we conjecture that this is the light asymptotic limit of the one-point function for all values of $\lambda_0$ and $z$ (up to a possible redefinition of $\lambda_0$), and not just in the limit (\ref{lzlim}). What we have rigorously established is however only the presence of the $\delta(\eta_1-\eta_2)\ (\Im z)^{-4\eta_1}$ factor, which follows from the $SL(3,\R)$ symmetry. This is already significant evidence that our classical and conformal bootstrap analyses of the $W-\bar{W}=0$ boundary condition actually describe the same D-branes.

We conclude with a few words on the boundary two-point function $\la e^{(\eta,\phi(x))}e^{(\mu,\phi(y))}\ra_{\lambda_0}^{light}$. The $SL(3,\R)$ symmetry determines its dependence on the boundary coordinates $x,y$, and implies that it must vanish unless the momenta of the two boundary operators are conjugate to each other, $\eta=\mu^*$. This relation between the two momenta can be confirmed by a direct calculation using the parametrization (\ref{dnt})-(\ref{xtwo}), which yields
\begin{multline}
\la e^{(\eta,\phi(0))}e^{(\mu,\phi(\infty))}\ra_{\lambda_0}^{light} = \delta^{(2)}(\eta-\mu^*) 
\int d^2a\ d^2b\ d^2c\ 
\\
\delta\left(\lambda_0-(\Im a^2+1)(\Im c^2+1)-(\Im b-\Re c\Im a)^2\right)\ \  (1+|b|^2+|c|^2)^{-\eta_1} (1+|a|^2+|ac-b|^2)^{-\eta_2}\ .
\label{eed}
\end{multline}
This integral is the same as $I_{\eta_1,\eta_2}$ (\ref{iee}) with an additional delta-function, and at this moment we do not know how to compute it. Also, we have no conformal bootstrap results to compare it with. In the special case $\lambda_0=-1$ when the boundary action is local (\ref{lz}), the exact result is known \cite{fat01}, but does not have a light asymptotic limit. Correspondingly, our light asymptotic calculation is meaningful only if $\lambda_0\geq 1$, because if $\lambda_0<1$ the argument of the $\delta$ function in eq. (\ref{eed}) cannot vanish.

\subsection{Case of $s\ell_3$ Toda theory with $W+\bar{W}=0$ \label{sslap}}

Let us consider the solutions (\ref{xpsol}) of the $s\ell_3$ Toda equations which obey the ``light asymptotic condition'' $\p^3X_i=\bp^3X_i=0$. We will write them in terms of the same functions $B(z)=(\frac12 z^2,z,1)$ as in the $W-\bar{W}=0$ case, together with the same matrix $P=\bsm 0 & 0 & 1 \\ 0 & -1 & 0 \\ 1 & 0 & 0\esm$ which relates them to their Wronskians, 
\bea
X_1=B(\bz) N P B(z)^T \scs X_2= B(\bz) P N^{-1T}  B(z)^T\ .
\eea
Now that we choose these particular functions $B(z)$, it is possible to ensure that $\phi_i=-\log X_i$ are real by imposing the simple condition that $NP$ be a positive Hermitian matrix. This restricts the $SL(3,\C)$ symmetry group (\ref{tpinv}) to its $SL(3,\R)$ subgroup, which still acts as $N\rar \Lambda^TN\Lambda^{-1T}$. (Beware that the matrix $P$ transforms nontrivially under this symmetry.) In addition, this implies that the eigenvalues of $N$ must be real, because there is a matrix $M$ such that $NP=\bar{M}^TM$, and $N$ is thus conjugate to the Hermitian matrix $MP\bar{M}^T$.

According to eq. (\ref{els}), the one-point function of a bulk operator $e^{(\eta,\phi(z))}=e^{\eta_1\phi_1(z)+\eta_2\phi_2(z)}$ in the presence of a boundary with parameters $\lambda_1,\lambda_2$ (\ref{ltn}) is of the type
\bea
\la e^{(\eta,\phi(z))} \ra_{\lambda_1,\lambda_2}^{light} =  \int dN\ \delta\left(\lambda_1-\Tr N\right)\ \delta\left(\lambda_2-\Tr N^{-1}\right)\ X_1^{-\eta_1}X_2^{-\eta_2}\ ,
\label{vpl}
\eea
The consequences of the $SL(3,\R)$ symmetry on this integral can be evaluated using the same particular symmetry transformations as in the case $W-\bar{W}=0$. Using $\Lambda(x,y)$ we similarly obtain the expected dependence on $z$. Using $\Lambda(d,e,f)$ however teaches us nothing new, because such transformations now leave $X_1,X_2$ invariant.  

Let us now use the parametrization $N=\bar{M}^TMP$ in terms of an upper-triangular matrix $M=\bsm \rho & a & b \\ 0 & \nu & c \\ 0 & 0 & \tau \esm$ which depends on three real parameters $\rho,\nu,\tau$ such that $\rho\nu\tau=1$, and three complex parameters $a,b,c$. The expressions (\ref{dnt})-(\ref{xtwo}) for $dN,X_1,X_2$ still hold, and we find
\bea
\lambda_1 &=& \Tr N = 2\rho\Re b-\nu^2-|a|^2\ ,
\\
\lambda_2 &=& \Tr N^{-1} = 2\rho\nu\Re(ac)-2\rho \nu^2 \Re b-\rho^2|c|^2 -\rho^2\tau^2\ .
\eea
We perform the shifts (\ref{shift}) in the expression (\ref{vpl}) for the one-point function, and obtain
\bea
\la e^{(\eta,\phi(z))} \ra_{\lambda_1,\lambda_2}^{light}&=&
\int dN\ \left(\tau^2+|b|^2+|c|^2\right)^{-\eta_1} \left(\rho^{-2}+\tau^2|a|^2+|\nu b-ca|^2\right)^{-\eta_2}
\nn \\
&& \times\
\delta\left(\lambda_1+\nu^2+|a|^2-2\rho\Re b+2\rho^2(\Im z)^2-4\Im a \Im z\right)\ 
\nn \\
&& \times\ 
\delta\left(\lambda_2+\nu^2\lambda_1+\nu^4+\nu^{-2}+|\rho c-\nu \bar{a}-2i\rho\nu\Im z|^2\right)
\ .
\label{vbzl}
\eea
We do not know how to perform this integral, except by taking a particular limit of the variables $z,\lambda_1,\lambda_2$. To this end, we parametrize the eigenvalues of $N$ as $\{e^{-(h,\sigma)}\}=\{ e^{-\sigma_1},e^{\sigma_1-\sigma_2},e^{\sigma_2}\}$ for a vector $\sigma=\sigma_1e_1+\sigma_2e_2$ in the Cartan subalgebra of $s\ell_3$. 
Then the parameters $\lambda_i$ coincide with values of the fundamental and antifundamental characters of $s\ell_3$, 
\bea
\lambda_1=\chi_{\om_2}(\sigma) \scs \lambda_2=\chi_{\om_1}(\sigma)\ .
\label{lclc}
\eea
Notice however that the matrix $N$ is not positive, and $\sigma_1,\sigma_2$ are not expected to be real numbers. Rather, we assume that $e^{\sigma_1}$ and $e^{\sigma_2}$ are negative real numbers. 
We now introduce the limit
\bea
\bla
z\rar \xi z \\ e^{\sigma_i}\rar \xi^2 e^{\sigma_i}\ela \scs \xi\rar \infty\ ,
\label{zslim}
\eea
so that $\lambda_i\rar \xi^2 e^{\sigma_i}$. In this limit, the delta-functions in the integral (\ref{vbzl}) fix the variables $\rho,\nu,\tau$ to values proportional to the eigenvalues of $N$, namely
\bea
\rho^2=-\tfrac12 \frac{1}{(\Im z)^2} e^{\sigma_2} \scs \nu^2=\tfrac12 e^{\sigma_1-\sigma_2} \scs \tau^2=-4(\Im z)^2 e^{-\sigma_1}\ .
\eea
And we find that the integral takes the value 
\bea
\la e^{(\eta,\phi(z))} \ra_{\sigma}^{light} &\sim& \frac{1}{\xi^4} (\Im z)^{-2\eta_1-2\eta_2} 
(-e^{\sigma_1})^{\eta_1-1} (-e^{\sigma_2})^{\eta_2-1}
I_{\eta_1,\eta_2}\ ,  \\
\frac{\la e^{(\eta,\phi(z))} \ra_{\sigma}^{light}}{\la 1 \ra_{\sigma}^{light}} &=& \frac{-2}{(\Im z)^{2(\rho,\eta)}} (-e^{\sigma_1})^{\eta_1} (-e^{\sigma_2})^{\eta_2}  \prod_{e>0}\frac{1}{(e,\eta-\rho)}\ ,
\label{eep}
\eea
where we used the definition (\ref{iee}) and value (\ref{ieer}) of the integral $I_{\eta_1,\eta_2}$. We derived this result in the case of $s\ell_3$ Toda theory, but it is not very difficult to generalize it to the case of $s\ell_n$ Toda theory with arbitrary $n$. 
By analogy with Liouville theory, we conjecture that this result holds for general values of $z$ and $\sigma$, and not just in the limit (\ref{zslim}). 

Let us investigate the light asymptotic limit (\ref{bes}) of the conformal bootstrap result (\ref{opt}). The behaviour of the factor $\sum_{w\in W} e^{(w(s),\al-Q)}$ depends on which Weyl chamber $s=b^{-1}\sigma$ belongs to; there is a Weyl chamber such that
\bea
\la V_{b\eta}(z)\ra_{b^{-1}\sigma} &\underset{b\rar 0}{\sim}& \frac{(\pi \mu b^{-2})^{-(\rho,\eta)}}{(2\Im z)^{2(\rho,\eta)} } e^{-b^{-2}(\rho,\sigma)} e^{(\eta-\rho,\sigma)}\  \prod_{e>0} \frac{\G(1+(e,\eta-\rho)-b^{-2}(e,\rho))}{(e,\eta-\rho)}\ ,
\\
\frac{\la V_{b\eta}(z)\ra_{b^{-1}\sigma}}{\la V_0(i)\ra_{b^{-1}\sigma}} &\underset{b\rar 0}{\sim}& -2
\frac{(\pi \mu b^{2})^{-(\rho,\eta)}}{(\Im z)^{2(\rho,\eta)} } (-1)^{2(\rho,\eta)} e^{(\eta,\sigma)} \prod_{e>0} \frac{1}{(e,\eta-\rho)}\ ,
\label{vol}
\eea
which generalizes the Liouville result (\ref{vllim}). 
Checking the agreement (\ref{vei}) between the classical calculation (\ref{eep}) and the conformal bootstrap result (\ref{vol}) involves assuming that the boundary parameters $e^{\sigma_i}$ which we introduced in both analyses coincide. Then the boundary cosmological constants $\lambda_{i,+}$ (\ref{lch}) agree with the classical boundary parameters $\lambda_{1,2}$ (\ref{lclc}).
However, as in Liouville theory, we must analytically continue the one-point function, because 
the boundary parameters $e^{\sigma_i}$ take positive values in the bootstrap analysis and negative values in the classical calculation. 

This agreement between the classical and bootstrap analyses, and the identification of their respective boundary parameters, have interesting consequences in the case when $\sigma$ belongs to the boundary of a Weyl chamber, that is $(e,\sigma)=0$ for $e$ some positive root. This is the case when two of the eigenvalues $\{ e^{-\sigma_1},e^{\sigma_1-\sigma_2},e^{\sigma_2}\}$ of the matrix $N$ coincide; in the classical analysis of Subsection \ref{ssp} this corresponded to the D-brane being one-dimensional. 
In the conformal bootstrap analysis, this case corresponds to the simply degenerate D-branes, as is clear from eq. (\ref{lpm}) for the boundary cosmological constants, where two of the three terms coincide. In the limit $b\rar 0$ with $c=-\frac13 b \kappa$ fixed and $m$ odd, the boundary cosmological constants become $\bla \lambda_{1,+}=e^{2c}-2e^{-c}\\ \lambda_{2,+}=e^{-2c}-2e^{c}\ela$. This allows us to identify $c$ with the position of the D-brane, as given by the Dirichlet conditions eq. (\ref{dnc}) from the classical analysis. 

We conclude with a few words on the boundary two-point function $\la e^{(\eta,\phi(x))}e^{(\mu,\phi(y))}\ra_{\lambda_1,\lambda_2}^{light}$. The $SL(3,\R)$ symmetry determines its dependence on the boundary coordinates $x,y$, and implies that it must vanish unless the momenta obey $(\rho,\eta-\mu)=0$. This condition can be interpreted as the conformal invariance of the boundary theory; however, assuming the momenta to be related to the $W^{(3)}$ charges as in eq. (\ref{qa}), we would expect a stronger constraint from the full $W_3$ symmetry, namely $\eta=\mu^*$. And indeed this is the constraint we found in the $W-\bar{W}=0$ case. But we can confirm the absence of this constraint in the present case by the explicit calculation
\begin{multline}
\la e^{(\eta,\phi(0))}e^{(\mu,\phi(\infty))}\ra_{\lambda_1,\lambda_2}^{light} =\delta((\rho,\eta-\mu)) 
\int d^2a\ d^2b\ d^2c\ \frac{d\nu}{\nu}\ \nu^{2(\eta_1-\mu_2)}
\\
\times \delta\left(\lambda_1-2\nu^{-1}\Re b+\nu^2+\nu^2|a|^2\right)\ 
\delta\left(\lambda_2+2\nu\Re b+\nu^{-2}+\nu^{-2}|c|^2-2\nu\Re(ac)\right)
\\
\times (1+|b|^2+|c|^2)^{-\eta_1} (1+|a|^2+|ac-b|^2)^{-\eta_2}
\ .
\label{bbbb}
\end{multline}
Curiously, the limit in which we are able to compute this integral is different from the limit (\ref{zslim}) which we used in the case of the one-point function, and in particular no longer forbids the coincidence of two eigenvalues of $N$. This limit is chosen so that a rescaling of $\nu$ can match the behaviours of $\lambda_1$ and $\lambda_2$ in the delta-functions: 
\bea
\bla
e^{\sigma_1}\rar \xi^{-1} e^{\sigma_1}\\ e^{\sigma_2}\rar \xi e^{\sigma_2}\ela \scs \xi\rar \infty\ .
\label{sslim}
\eea
We obtain in this limit
\begin{multline}
\xi^{3+2\eta_2-2\mu_1}\la e^{(\eta,\phi(0))}e^{(\mu,\phi(\infty))}\ra_{\sigma}^{light} \sim \delta((\rho,\eta-\mu)) 
\frac{1}{(\eta_2-1)(\mu_2-1)} \frac{\G(\eta_1+\eta_2-\frac32)}{\G(\eta_1+\eta_2-1)}
\\ \times |e^{\sigma_2-\sigma_1}|^{\mu_1-\eta_2-\frac32} \left|\sinh \tfrac12(\sigma_1+\sigma_2)\right|^{-2\eta_1-2\eta_2+3}\ .
\label{eegg}
\end{multline}
This result is invariant under the exchange of the two operators $\eta\lrar \mu$. On the other hand there is no invariance under $\eta\rar \eta^*,\mu\rar \mu^*$. This is because our limit (\ref{sslim}) treats the boundary parameters $\lambda_1,\lambda_2$ in an asymmetric way; in particular $\lambda_1$ and $\lambda_2$ do not go to infinity at the same rate.

\zeq\section{Conclusion \label{secco}}

Combining classical and conformal bootstrap analyses yields a consistent picture of the moduli space of maximally symmetric D-branes in $s\ell_n$ conformal Toda theory.
Our results and conjectures on these moduli spaces and on the existence of a boundary action in the $s\ell_2$ and $s\ell_3$ cases can be summarized in the following table:
\begin{center}
\begin{tabular}{|l|l|c|c|c|c|}
\hline
Theory & Type of Brane & $d$ & Parameters & \!\!Classical parameters\!\! & Action
\\
\hline
\hline
\multirow{2}{*}{Liouville} & Continuous & $1$ & $s\in \R$ & $\lambda_L= i\cosh bs$ & $\lambda_L\int e^{\phi}$ 
\\  & Discrete & $0$ & $\ell|\ell'\in \N^2$ &  n. a. & n. a. 
\\
\hline
\multirow{2}{*}{$s\ell_3$, $W=\bar{W}$} & Continuous & $2$ &  ? & $\lambda_0$ & nonlocal
\\ & Degenerate & ? & ? & ? & ?
\\
\hline
\multirow{3}{*}{$s\ell_3$, $W=-\bar{W}$} & Continuous & $2$ & $s\in \R^2$ & $\lambda_i=\chi_{\om_i}(bs)$ & inexistent 
\\ & Simply degenerate\! & $1$ & $\kappa|\ell,m\in \R\times \N^2$ & $c=-\tfrac13 b\kappa$ & $0$ 
\\ & Discrete & $0$ & $\ell,m|\ell',m'\in \N^4$ & n. a. & n. a.
\\
\hline
\end{tabular}
\end{center}

In the case of the boundary conditions $W^{(s)}-\bar{W}^{(s)}=0$, the dimension of the moduli space is the integer part of $\frac{n}{2}$ (Subsection \ref{ssm}). This coincides with the number of nonzero charges $\{q^{(s)}\}_{2\leq s\leq n-1|s\ {\rm even}}$ for bulk operators whose one-point functions do not vanish (Section \ref{secbo}), in accordance with a generalization of Cardy's idea. In the $s\ell_3$ case, we provide predictions for certain correlation functions in the light asymptotic limit, namely the bulk one-point function (\ref{eet}) and boundary two-point function (\ref{eed}).

We examined the case of the boundary conditions $W^{(s)}-(-1)^s \bar{W}^{(s)}=0$ in more detail. We propose that there exists a hierarchy of D-branes of dimensions $d=0\cdots n-1$, which correspond to  
representations of the $W_n$ algebra with $\frac12 d(d+1)$ null vectors. In particular, there are continuous D-branes of dimension $n-1$, and discrete D-branes of dimension $0$. The moduli space of $d$-dimensional D-branes is itself $d$-dimensional, although there are also discrete parameters. This was the result of classical (Subsection \ref{ssp}) and bootstrap (Section \ref{secbo}) analyses, which were shown to agree in detail (Subsection \ref{sslap}). In particular, we found explicit formulas for the bulk one-point functions of continuous (\ref{opt}) and discrete (\ref{uww}) D-branes. In the $s\ell_3$ case, we also computed the bulk one-point functions of the simply degenerate D-branes (\ref{uess}). As our D-branes conform to Cardy's ideas by corresponding to representations of the $W_n$ algebra, they also correspond to the topological defects of the very interesting article \cite{dgg10} (where such defects are related to certain operators in four-dimensional gauge theories). And a D-brane's one-point function is closely related to the corresponding defect operator's coefficients. 

The calculation of annulus partition functions leads to natural conjectures for the spectra of open strings with one end on a discrete D-brane (Subsection \ref{ssmo}). These spectra coincide with what can be obtained by fusing the two representations which correspond to the two involved D-branes. If this structural property persists in the case of all D-branes, then it can help explain the divergences of the annulus partition functions (Subsection \ref{ssspec}). Infinite fusion multiplicities indeed appear in the fusion of two continuous representations (Subsection \ref{ssmul}), so that we expect infinite multiplicities in the spectra of continuous D-branes. This might also explain the apparent violation of the $W_3$ symmetry in the minisuperspace prediction (\ref{eegg}) for the boundary two-point function. A boundary spectrum with infinite multiplicities can certainly not be adequately parametrized by momenta $\al=b\eta$, and an operator with a given momentum might correspond to a combination of states belonging to different representations of the $W_3$ algebra. 

Thus, the moduli space of D-branes may now be well-understood, but the boundary operators and their correlation functions remain problematic, and they certainly have new and complicated features.

\acknowledgments{We are grateful to Alexei Litvinov and Philippe Roche for interesting discussions. Moreover, we wish to thank Philippe Roche and Volker Schomerus for comments on the draft of this article. This work was supported in part by the cooperative CNRS-RFBR grant
PICS-09-02-93106.
}

\appendix

\zeq\section{Minisuperspace limits of some correlation functions \label{secmi}}

In addition to the light asymptotic limit which we studied in Section \ref{secla}, there is another semi-classical limit in which Toda correlation functions simplify and can in certain cases be independently predicted: the minisuperspace limit, where our two-dimensional field theory reduces to a one-dimensional system. In this limit, a bulk primary operator $V_{Q+ip}(z)$ corresponds to a wavefunction $\Psi_p(\phi)$, which is a solution of the
Schr\"odinger equation of Toda quantum mechanics \cite{fl07c},
\bea
\left[-\left(\pp{\phi}\right)^2+2\pi \mu b^{-2} \sum_{i=1}^{n-1} e^{(e_i,\phi)}\right] \Psi_p(\phi) = p^2 \Psi_p(\phi)\ .
\label{sch}
\eea
(Compare with the  $s\ell_n$ Toda Lagrangian $L_n$ (\ref{ln}), and notice the rescaling $\phi\rar b^{-1}\phi$.) Here the variable $\phi$ can be interpreted as the $z$-independent zero-mode of the Toda field $\phi(z)$. The Schr\"odinger equation is deduced from the Hamiltonian picture of the dynamics of $\phi$, which is associated to radial quantization in the $z$-plane.

In the minisuperspace limit, a boundary with parameter $\sigma$ corresponds to a boundary wavefunction $\Psi_\sigma^{bdy}(\phi)$, which can be interpreted as the density of the corresponding D-brane. If a boundary Lagrangian $L^{bdy}[\phi]$ is known, then the boundary wavefunction can be obtained by computing this Lagrangian for constant values of the field $\phi(z)$, namely 
\bea
\Psi_\sigma^{bdy}(\phi)=e^{-L^{bdy}(\phi)}\ .
\label{pel}
\eea
In any case, the minisuperspace one-point function is defined as 
\bea
\la \Psi_p \ra_{\sigma}^{mini} \equiv  \int d\phi \ \Psi_p(\phi)\ \Psi_\sigma^{bdy}(\phi) \ ,
\label{pipp}
\eea
and we expect that it is related to a $b\rar 0$ limit of the one-point function $\la V_\al(z) \ra_s$,
\bea
\underset{b\rar 0}{\lim} (\Im z)^{2\Delta_{Q+ibp}} \la V_{Q+ibp}(z) \ra_{b^{-1}\sigma} = \la \Psi_p \ra_{\sigma}^{mini}\ .
\label{lvp}
\eea
(Compare with the light asymptotic limit (\ref{vei}).) 

In the case of Liouville theory, a boundary Lagrangian is known. Then it is possible to compute the minisuperspace one-point function (\ref{pipp}) and to compare it with the conformal bootstrap one-point function. It turns out that eq. (\ref{lvp}) is obeyed, which provides a test of the conformal bootstrap one-point function \cite{fzz00}. In the case of $s\ell_3$ Toda theory with $W+\bar{W}=0$, no boundary action exists, as we will see in Appendix \ref{secac}. We will reason in the opposite direction, and deduce the minisuperspace boundary wavefunction $\Psi_\sigma^{bdy}(\phi)$ from the conformal bootstrap one-point function. We will do this first in Liouville theory, as a preparation for the case of $s\ell_3$ Toda theory. 
The boundary wavefunction will turn out to have interesting properties; in particular it provides the generating function of the B\"acklund transformation which maps 
the Toda classical mechanics of $\phi$ to the free classical mechanics of $\sigma$. As we saw in Section \ref{sted}, there is a good reason why the Toda boundary parameter $\sigma$ can be interpreted as a free field: there exists a B\"acklund transformation from conformal Toda theory to a free field theory, such that the $W+\bar{W}=0$ boundary conditions in Toda theory are mapped to Dirichlet boundary conditions in the free theory, and the Toda boundary parameter $\sigma$ is mapped to the free field boundary parameter, which is the boundary value of the free field.

\subsection{Case of Liouville theory}

In this case the Schr\"odinger equation (\ref{sch}) becomes $\left[-\frac12\ppd{\phi}+2\pi \mu b^{-2} e^{2\phi}\right] \Psi_p(\phi) = \frac12 p^2 \Psi_p(\phi)$. The solution is \cite{fzz00}
\bea
\Psi_p(\phi)=(\pi \mu b^{-2})^{-\frac{i}{2} p} \frac{2}{\G(-ip)} K_{ip}\left(2\sqrt{\pi \mu b^{-2}} e^{\phi}\right)\ ,
\eea
where $K_\nu(z)$ is a Bessel function, and $\Psi_p$ is normalized so that
\bea
\int d\phi\ \Psi_{p_1}(\phi) \Psi_{p_2}(\phi)=2\pi \delta(p_1+p_2)\ .
\eea
The minisuperspace limit (\ref{lvp}) of the Liouville one-point function (\ref{opt}) is 
\bea
\la \Psi_p\ra^{mini}_\sigma = 2(\pi \mu b^{-2})^{-\frac{i}{2} p} \G(ip) \cos (p \sigma)\ .
\eea
According to eq. (\ref{pel}) and eq. (\ref{pipp}) we can deduce the boundary wavefunction from the knowledge of $\la \Psi_p\ra^{mini}_\sigma$,
\bea
\Psi^{bdy}_\sigma(\phi) = \frac{1}{2\pi} \int dp\ \la \Psi_{-p}\ra^{mini}_\sigma \Psi_{p}(\phi) = \frac{2}{\pi} \int dp\ \cos (p \sigma)\ K_{ip}\left(2\sqrt{\pi \mu b^{-2}} e^{\phi}\right)\ .
\eea
The calculation is performed using the formula $\int_0^\infty dp\ \cos (ap)\ K_{ip}(z)=\frac{\pi}{2} e^{-z\cosh a}$, with the result
\bea
\Psi^{bdy}_\sigma(\phi) = e^{-L_2^{bdy}}\scs
L_2^{bdy}=\sqrt{4\pi \mu b^{-2}} \cosh(\sigma)\ e^{\phi}\ .
\eea
The function $L_2^{bdy}(\phi,\sigma)$ generates a canonical transformation between Toda and free classical mechanics, as follows from the identity
\bea
 \left(\frac{\p L_2^{bdy}}{\p \phi}\right)^2- \left(\frac{\p L_2^{bdy}}{\p\sigma}\right)^2=4\pi \mu b^{-2}\ e^{2\phi}\ , 
\label{pplv}
\eea
whose right hand-side is the bulk Liouville potential, see the Lagrangian (\ref{ln}). Considering indeed $\phi,\sigma$ as time-dependent variables with associated momenta $\dot{\phi}=-\frac{\p L_2^{bdy}}{\p \phi}$ and $\dot{\sigma}=\frac{\p L_2^{bdy}}{\p\sigma}$, the Liouville equation of motion $\ddot{\phi}=4\pi\mu b^{-2}\ e^{2\phi}$ amounts to the $\phi$-derivative of eq. (\ref{pplv}), and the free equation of motion $\ddot{\sigma}=0$ amounts to the $\sigma$-derivative of eq. (\ref{pplv}).

\subsection{Case of $s\ell_3$ Toda theory with $W+\bar{W}=0$}

The solution of the Schr\"odinger equation (\ref{sch}) in the case of $s\ell_3$ Toda theory is \cite{tv78} \cite{fl07c}
\begin{multline}
\Psi_p(\phi)=\frac{8(\pi \mu b^{-2})^{-i(\rho,p)}}{\prod_{e>0}\G(-i(e,p))} e^{\frac{i}{6}(e_1-e_2,\phi)(e_1-e_2,p)} 
\\ \times 
\int_0^\infty\frac{dt}{t} t^{i(e_2-e_1,p)} K_{i(\rho,p)}\left(2\sqrt{1+t^{-2}} \sqrt{\frac{\pi\mu}{b^{2}}} e^{\frac12(e_1,x)}\right)
K_{i(\rho,p)}\left(2\sqrt{1+t^2} \sqrt{\frac{\pi\mu}{ b^{2}}} e^{\frac12(e_2,x)}\right)\ ,
\end{multline}
and it is normalized such that
\bea
\int d^2\phi\ \Psi_{p_1}(\phi)\Psi_{p_2}(\phi) = (2\pi)^2 \delta^{(2)}(p_1+p_2)\ .
\eea
The minisuperspace limit (\ref{lvp}) of the $s\ell_3$ Toda one-point function (\ref{opt}) is 
\bea
\la \Psi_p\ra^{mini}_\sigma=(\pi \mu b^{-2})^{-i(\rho,p)} \prod_{e>0} \G(i(e,p)) \sum_{w\in W} e^{i(w(\sigma),p)}\ .
\eea
According to eq. (\ref{pel}) and eq. (\ref{pipp}) we can deduce the boundary wavefunction from the knowledge of $\la \Psi_p\ra^{mini}_\sigma$,
\bea
\Psi^{bdy}_\sigma(\phi) = \frac{1}{(2\pi)^2} \int d^2p\ \la \Psi_{-p}\ra^{mini}_\sigma \Psi_{p}(\phi) \ .
\eea
The calculation can be performed using the formula 
\bea
\int_0^\infty dp\ \cos(p\sigma)\ K_{ip}(z_1) K_{ip}(z_2) = \frac{\pi}{2} K_0\left(\sqrt{z_1^2+z_2^2+2bc \cosh \sigma}\right)
\ .
\eea
The result is 
\bea
\Psi^{bdy}_\sigma(\phi) =  K_0\left(L_3^{(0)}\right)\ ,
\label{elk}
\eea
where we define
\bea
L_3^{(0)}&\equiv& \sqrt{4\pi\mu b^{-2}}\sqrt{ e^{(e_1,\phi)}+e^{(e_2,\phi)}+e^{(\om_1,\phi)}\chi_{\om_1}(\sigma^*) +e^{(\om_2,\phi)} \chi_{\om_2}(\sigma^*)}\ ,
\\
&=& \sqrt{4\pi\mu b^{-2}} \prod_{h\in H_{\om_1}}\sqrt{e^{\frac13(e_1,\phi)}+(-1)^{(\rho,h)} e^{\frac13(e_2,\phi)}e^{(h,\sigma^*)}}\ .
\eea
In the strong coupling region where $\mu$ is large, we have $\Psi^{bdy}_\sigma(\phi)\sim e^{-L_3^{(0)}}$ as follows 
from $K_0(z)\underset{z\rar\infty}{\sim} \sqrt{\frac{\pi}{2z}} e^{-z}$. And $L_3^{(0)}$ generates the canonical transformation from Toda classical mechanics to the free classical mechanics. (The transformation itself is written in \cite{anps94}.)
As in the case of Liouville theory, this follows from the identity
\bea
\left(\frac{\p L_3^{(0)}}{\p \phi}\right)^2- \left(\frac{\p L_3^{(0)}}{\p\sigma}\right)^2=4\pi \mu b^{-2} \left[e^{(e_1,\phi)}+e^{(e_2,\phi)}\right]\ ,
\label{pplt}
\eea
which can be proved with the help of the formulas 
\bea
\bla
\left(\frac{\p \chi_{\om_1}(\sigma)}{\p \sigma}\right)^2=\tfrac23 \chi_{2\om_1}(\sigma)-\tfrac43 \chi_{\om_2}(\sigma)
\\
\left(\frac{\p \chi_{\om_1}(\sigma)}{\p \sigma},\frac{\p \chi_{\om_2}(\sigma)}{\p \sigma}\right) =\tfrac13\chi_\rho(\sigma) -\tfrac83
\ela
\scs
\bla
\chi_{\om_1}^2=\chi_{2\om_1}+\chi_{\om_2}
\\
\chi_{\om_1}\chi_{\om_2}=\chi_{\rho}+1
\ela
\ \ \ .
\eea

To conclude, let us come back to the interpretation of the boundary wavefunction $\Psi^{bdy}_\sigma(\phi)$ as the density of the continuous D-brane of parameter $\sigma$, as suggested by eq. (\ref{pipp}).
In the weak coupling region where $L_3^{(0)}$ is small, we have $ \Psi^{bdy}_\sigma(\phi) \underset{(\rho,\phi)\rar -\infty}{\sim} -(\rho,\phi)$ as follows from $K_0(z)\underset{z\rar 0}{\sim} -\log\frac{z}{2}$. Thus, the density of the D-brane grows linearly with $\phi$. The minisuperspace annulus partition function $Z_{\sigma_1;\sigma_2}^{mini} = \int d^2\phi\ \Psi_{\sigma_1}(\phi)\Psi_{\sigma_2}(\phi)$ therefore has an $L^4$ infrared divergence, where $L$ is a large distance cutoff. This confirms the divergence which was found by modular bootstrap methods in Subsection \ref{ssmo}. This contrasts with the case of Liouville theory, where the density of a continuous D-brane is constant in the weak coupling region, and correspondingly the annulus partition function diverges as $L$, which is the volume of the $\phi$-space in that case.

\zeq\section{On the existence of a boundary action in $s\ell_3$ Toda theory \label{secac}}

The functional integral formalism is often useful in the study of conformal field theories, although in general it permits the calculation of only a subset of the correlation functions. In this formalism, correlation functions are expressed as functional integrals over the fields $\phi_i$, where field configurations come with weights $e^{-S}$. Here $S$ is the action, which may or may not be written as the integral of a certain Lagrangian $L$, namely $S=\int d^2z\ L(z)$. If the Lagrangian exists and is local, that is if $L(z)$ is a function of the fields $\phi_i$ and finitely many of their derivatives at the point $z$, then the action is also called local. If the space has a boundary $z=\bz$, the boundary action or boundary terms of the action are the terms which depend only on the values of the fields at the boundary, and local boundary actions are those of the type $S=\int_{z=\bz} dx\ L(x)$ where $L(x)$ is a local boundary Lagrangian.

The choice of an action $S$ is constrained by the classical theory. Namely, the solutions of the classical equations of motion and boundary conditions should be functional critical points of the action. This constraint does not fully determine $S$; here we will however only be concerned with the question of the existence of at least one action which obeys this constraint.

In $s\ell_3$ conformal Toda theory on surfaces with no boundaries, the Lagrangian $L_3$ (\ref{ln}) is known \cite{fl07c}. In Liouville theory on surfaces with boundaries, we have the boundary Lagrangian $L_2^{bdy}$ (\ref{ltb}), see \cite{fzz00}. In the case of $s\ell_3$ Toda theory, we could so far derive our boundary conditions from  boundary Lagrangians only in particular subcases of the two cases $W=\pm\bar{W}$. 
We will now investigate systematically for which boundary conditions ($W=\pm\bar{W}$) and boundary parameters ($\lambda_0$ or $\lambda_1,\lambda_2$) boundary actions can exist.

\subsection{Boundary conditions as functional one-forms}

Let us assume the existence of a boundary action $S^{bdy}[\phi_i]$, that is a functional of the values of the Toda fields $\phi_1,\phi_2$ at the boundary. We however do not assume that $S^{bdy}[\phi_i]$
is local. In particular we do not forbid introducing auxiliary boundary fields in addition to $\phi_i$, so long as these auxiliary fields can be eliminated using their equations of motion. We only exclude the possibility for fields to obey Dirichlet boundary conditions, which excludes the particular case (\ref{dnc}) from the analysis.

Let us derive the Neumann-type boundary conditions from the action $S=\int d^2z \frac12 (\p\phi,\bp\phi) + S^{bdy}_3[\phi_i]$, where the interaction terms in the bulk action (\ref{ln}) can be omitted as they will not contribute. We find 
\bea
\frac{1}{2i}(\p-\bp)(2\phi_1-\phi_2)=\frac{\delta S^{bdy}}{\delta \phi_1} \scs \frac{1}{2i}(\p-\bp)(2\phi_2-\phi_1)=\frac{\delta S^{bdy}}{\delta \phi_2}\ .
\eea
In terms of the $X_i=e^{-\phi_i}$, this becomes
\bea
2\frac{(\p-\bp)X_1}{X_1^2}-\frac{(\p-\bp)X_2}{X_1X_2} =2i\frac{\delta S^{bdy}}{\delta X_1} \scs 
2\frac{(\p-\bp)X_2}{X_2^2}-\frac{(\p-\bp)X_1}{X_1X_2} =2i\frac{\delta S^{bdy}}{\delta X_2}\ .
\label{pxsb}
\eea
The existence of the boundary action $S^{bdy}$ can now be interpreted as the condition that the
functional one-form
\bea
g=\left(2\frac{(\p-\bp)X_1}{X_1^2}-\frac{(\p-\bp)X_2}{X_1X_2}\right)\delta X_1 + \left(2\frac{(\p-\bp)X_2}{X_2^2}-\frac{(\p-\bp)X_1}{X_1X_2}\right)\delta X_2\ 
\label{gform}
\eea
be exact, namely $g=\delta(2iS^{bdy})$. It follows that $g$ must be closed, $\delta g=0$. In order to be able to work with this condition, we will study functional calculus in the next Subsection.

Before that, let us point out that the natural variables to work with are not $X_1,X_2$ but the functions $b_1,b_2,b_3$ in terms of which we wrote the solutions of the Toda equations (\ref{xmsol}) and (\ref{xpsol}). These variables are subject to the constraint ${\rm Wr}[b_1,b_2,b_3]=1$, so that we must include the possibility of such constraints in our study of functional calculus.

\subsection{Technical interlude: functional calculus}

We wish to study functional forms which depend on functions $b_i(x)$. A zero-form is a functional $S[b_i]$. A one-form is an object $g=\int dx \sum_i g_i(x) \delta b_i(x)$, where $g_i(x)$ are $x$-dependent functionals of $b_i$. An example of a one-form is the differential of a zero-form, namely $\delta S = \int dx\ \sum_i \frac{\delta S}{\delta b_i(x)} \delta b_i(x)$. A two-form is an object $k=\int dxdy \sum_{ij}k_{ij}(x,y) \delta b_i(x)\wedge \delta b_j(y)$, where $k_{ij}(x,y)$ are $x,y$-dependent functionals of $b_i$. The basic two-forms $\delta b_i(x)\wedge \delta b_j(y)=-\delta b_j(y)\wedge \delta b_i(x)$ are antisymmetric, which however does not imply the vanishing of $\delta b_i(x)\wedge \delta b_i(y) = -\delta b_i(y)\wedge \delta b_i(x)$. So the differential of a one-form is
\begin{multline}
\delta\left(\sum_i\int dx\ g_i(x)\delta b_i(x)\right)=\sum_{i<j}\int dxdy \left(\frac{\delta g_i(x)}{\delta b_j(y)}-\frac{\delta g_j(y)}{\delta b_i(x)}\right) \delta b_i(x)\wedge \delta b_j(y) 
\\ +\sum_i\int dxdy \frac{\delta g_i(x)}{\delta b_i(y)} \delta b_i(x)\wedge \delta b_i(y)\ .
\end{multline}
As an exercise, we can compute the differential of an action functional $S=\int dx\ L(b(x),b'(x))$,
\bea
\delta S = \int dzdy \left[\frac{\p L}{\p b}(z) \delta(z-y)+\frac{\p L}{\p b'}(z) \delta'(z-y)\right]\!\delta b(y)=\int dz \left[\frac{\p L}{\p b}-\pp{z} \frac{\p L}{\p b'}\right]\!\!(z) \delta b(z)\ ,
\eea 
and we can check that $\delta^2 S=0$. 

Now we will be interested in variables $b_1,b_2,b_3$ which are not independent, as they obey the constraint ${\rm Wr}[b_1,b_2,b_3]=1$. If these were ordinary variables instead of functions, the condition for the form $g=\sum g_i db_i$ to be closed modulo a constraint $C(b_1,b_2,b_3)=1$ would simply be $dg \wedge dC=0$, and the integral $S$ of the one-form $g$ would be characterized by $(dS-g)\wedge dC=0$. Let us generalize these notions to the case of functional forms. Let $g=\int dx \sum_{i=1}^3 g_i(x) \delta b_i(x)$ be a one-form, let us study the condition that it is closed modulo the constraint ${\rm Wr}$. We denote $\delta g=\int dxdy \sum_{ij}k_{ij}(x,y) \delta b_i(x)\wedge \delta b_j(y)$ with $k_{ij}(x,y)=-k_{ji}(y,x)$. 

We assume for a moment that the constraint can be inverted and rewritten as $b_3=\phi[b_1,b_2]$. Then it is straightforward to rewrite $g=\int dx \sum_{i=1}^2 \tilde{g}_i(x)\delta b_i(x)$ and to compute $\delta g$ in terms of $\phi$ and $k_{ij}$. We find that the vanishing of $\delta g$ modulo the constraint ${\rm Wr}[b_1,b_2,b_3]=1$ is equivalent to
\bea
K_{12}-K_{13}-K_{32}+K_{33} = K_{11}-K_{13}-K_{31}+K_{33}=K_{22}-K_{23}-K_{32}+K_{33}=0\ ,\quad &&
\label{keq}
\\ {\rm where} \ \ \ K_{ij}\equiv \left(\frac{\delta {\rm Wr}}{\delta b_i}\right)^{-1t} k_{ij} \left(\frac{\delta {\rm Wr}}{\delta b_j}\right)^{-1}\ .\quad &&
\label{kij}
\eea
In the definition of $K_{ij}$ we have used new notations for functions of two variables $f(x,y)$ such as $\frac{\delta {\rm Wr}(x)}{\delta b_i(y)}$ or $k_{ij}(x,y)$. Namely, the products and inverses of such functions are defined with respect to the product law $(f_1f_2)(x,y)\equiv\int dz\ f_1(x,z)f_2(z,y)$, and the transposition is defined as the exchange of the two variables, $f^t(x,y)\equiv f(y,x)$. In the case when the functions $b_i$ are $x$-independent, the product law becomes commutative, the objects $k_{ii}$ and $K_{ii}$ vanish, and the conditions (\ref{keq}) boil down to $K_{12}+K_{23}+K_{31}=0$ which is equivalent to $dg\wedge d{\rm Wr}=0$ as we found by the direct analysis of that case. 
Notice that the conditions (\ref{keq}) on the matrix $K_{ij}$ are equivalent to $\sum_{ij} v_i K_{ij} v'_j=0$ for any two vectors $v,v'$ such that $\sum_i v_i=\sum_i v'_j=0$. 

Then the conditions for an ``action'' functional $S$ to be the integral of the functional one-form $g$ modulo the constraint ${\rm Wr}$ is:
\bea
\left(\frac{\delta S}{\delta b_1}-g_1\right) \left(\frac{\delta {\rm Wr}}{\delta b_1}\right)^{-1} = 
\left(\frac{\delta S}{\delta b_2}-g_2\right) \left(\frac{\delta {\rm Wr}}{\delta b_2}\right)^{-1} =
\left(\frac{\delta S}{\delta b_3}-g_3\right) \left(\frac{\delta {\rm Wr}}{\delta b_3}\right)^{-1} \ .
\label{sgwi}
\eea

Now the Wronskian constraint is not invertible, as $b_3$ cannot be fully determined in terms of $b_1,b_2$. So the quantities $\left(\frac{\delta {\rm Wr}}{\delta b_i}\right)^{-1}$ are ambiguous. We indeed find that $\frac{\delta {\rm Wr}}{\delta b_i}$ has several inverses, parametrized by numbers $\kappa^i_{jk}$,
\bea
\left(\frac{\delta {\rm Wr}}{\delta b_i}\right)^{-1}(x,y)= \frac{1}{w_i^2}\left[\Theta(y-x)\e_{ijk}b_j(x)b_k(y) +\sum_{j,k\neq i} \kappa^i_{jk} b_j(x)b_k(y)\right]\ ,
\label{wbm}
\eea
where $\Theta(x)$ is a step function such that $\Theta'(x)=\delta(x)$, and we recall the notations $w_i=\e_{ijk}b_jb'_k$ and ${\rm Wr}=\e_{ijk}b_ib'_jb''_k$. Then for $\delta g$ to vanish modulo the constraint, the condition (\ref{keq}) must hold for all values of $\kappa^i_{jk}$. Similarly, integrating the functional one-form $g$ modulo the constraint requires the equation (\ref{sgwi}) to be satisfied for all values of $\kappa^i_{jk}$. 

\subsection{Existence of the boundary action if $W-\bar{W}=0$ \label{bam}}

We have found that the boundary conditions $W-\bar{W}=0$ lead to the expressions (\ref{xmsol}) for the Toda fields $X_1,X_2$ in terms of functions $b_i$ subject to the Wronskian constraint. The expressions (\ref{dxm}) for $(\p-\bp)X_i$ are also known. These expressions depend on a constant matrix $N_{ij}$ of size $3$ and determinant $1$; it will be convenient to decompose both $N$ and $N^{-1T}$ into symmetric and antisymmetric parts, according to $N_{ij}=S_{ij}+\e_{ijk}A_k$ and $N^{-1T}=\sigma_{ij}+\e_{ijk}\al_k$. 

So we can compute the one-form $g$ (\ref{gform}) and its differential $k=\delta g$ in terms of the functions $b_i$. 
Taking the ambiguities $\kappa^i_{jk}$ to vanish in the inversion (\ref{wbm}) of the Wronskian constraint, the 
quantities $K_{ij}$ (\ref{kij}) turn out to be of the form 
\bea
K_{ij}(x,y)=\frac{2 \e_{i\ell m} \e_{jpq}b_\ell(x) b_p(y)}{w_i^2(x) w_j^2(y)} \int dz\ 
\Theta(x-z)\Theta(y-z)\ \  \Lambda^{ij}_{mq}(z)\ ,
\label{kijxy}
\eea
where we sum over repeated indices except $i,j$, and 
the tensor $\Lambda_{mq}^{ij}$, which is defined for $q\neq j$ and $m\neq i$ and obeys $\Lambda^{ij}_{mq}=-\Lambda^{ji}_{qm}$, is 
\bea
\Lambda^{ij}_{mq}  & = & 
\frac{2\sigma_{qr}w_r w_jb_m }{X_1X_2^2}(\al_i S_{uv}-\al_u S_{iv})b_ub_v 
-\frac{2\sigma_{mr}w_r w_i b_q}{X_1X_2^2}(\al_jS_{uv}-\al_uS_{jv})b_ub_v
\nn \\
& + & \frac{2S_{ir}b_r w_jb_m}{X_1^2X_2}(A_u\sigma_{qv}-A_q\sigma_{uv})w_uw_v 
-\frac{2S_{jr}b_r w_ib_q}{X_1^2X_2}(A_u\sigma_{mv}-A_m\sigma_{uv})w_uw_v
\nn \\ 
&+ & (\al_iS_{jr}-\al_jS_{ir}) \frac{b_rb_mb_q}{X_1X_2} +(A_m\sigma_{qr}-A_q\sigma_{mr}) \frac{w_iw_jw_r}{X_1X_2}\ .
\eea
In the special case of the free boundary conditions we have $A_m=\al_m=0$ thus $\Lambda^{ij}_{mq}=0$.
In the special case when the boundary Lagrangian given by eq. (\ref{lz}) we have $S_{ij}=U_iU_j$ thus $\al_i S_{uv}-\al_u S_{iv}=A_u\sigma_{qv}-A_q\sigma_{uv}=0$ thus again $\Lambda^{ij}_{mq}=0$.

For a quantity $K_{ij}$ of the form (\ref{kijxy}), the condition (\ref{keq}) amounts to
\bea \left\{\begin{array}{l} \forall m\neq i \\ \forall q\neq j\end{array}\right.,\
\int \Lambda^{ij}_{mq}=0\ \ {\rm and}\ \
\left\{\begin{array}{l} \forall j,k \\ \forall i\neq m\end{array}\right., \ \e_{jpq} \frac{b_p(y)}{w_j^2(y)}\int^y \Lambda^{ij}_{mq} =\e_{kpq}\frac{b_p(y)}{w_k^2(y)}\int^y \Lambda^{ik}_{mq}\ ,
\label{fjkim}
\eea
where $\int^y \Lambda$ is the primitive of the function $\Lambda$. Curiously, taking into account the ambiguities parametrized by $\kappa^i_{jk}$ does not yield extra equations. 

We wish to find out whether the equation (\ref{fjkim}) holds for any triples $(b_1,b_2,b_3)$ obeying the Wronskian constraint. We do not know how to do this except by testing the equation for a number of triples. Large families of solutions of the Wronskian constraint can be built from functions of the type $z^\nu$ or $e^{\nu z}$. This raises the questions of the admissible behaviour of $b_i(z)$ at $z=\infty$ and at generic points $z$, and of the appropriate contours of integration in our equation (\ref{fjkim}). We have no satisfactory answers to these questions. So we will test only the purely algebraic consequences of our equation. 

Consider an equation of the type $u_0(y)=0$ where $u_0(y)=\sum_{i=1}^n u_i(y)\int^y v_i$. Let us build the matrix of size $n+1$ formed by $u_i$ and their first $n$ derivatives, $M=[u_i^{(j)}]_{i,j=0\cdots n}$. Then $\det M=0$ is a purely algebraic consequence of the original equation, in the sense that the terms involving primitives $\int^y v_i $ cancel. Applying this treatment to eq. (\ref{fjkim}) removes the need to deal with integrals and to worry about the regularity of $b_i(z)$. For all the numerous cases which we tested, we found that the condition $\det M=0$ held. This is strong evidence that the form $\delta g$ is closed. This is strong evidence that it is in fact exact, and we conjecture that there exists a boundary action from which the boundary condition (\ref{anb}) can be derived. 

This action is expected to be a functional $S^{bdy}$ of the values of the Toda fields $\phi_1,\phi_2$ at the boundary $z=\bz$. In addition, $S^{bdy}$ is expected to depend on the boundary parameter $\lambda_0$. Comparing its definition (\ref{pxsb}) with the formulas (\ref{xmsol}) for $X_i$ and (\ref{dxm}) for $(\p-\bp)X_i$, we see that $S^{bdy}$ cannot be local, that is of the type $\int L^{bdy}[\phi_1,\phi_2]$ where $L^{bdy}$ is a function of $\phi_i$ and finitely many of their derivatives, except in the two special cases $\lambda_0=\pm 1$ which we considered in Subsection \ref{ssm}. 
It is possible that the nonlocal boundary action has a simple expression as a local functional of $b_i$. Even so, it would not be very easy to use such an action in free-field computations of correlation functions.

\subsection{No boundary action if $W+\bar{W}=0$ \label{bap}}

We have found that the boundary condition $W+\bar{W}=0$ led to the expressions (\ref{xpsol}) for the Toda fields $X_1,X_2$ in terms of functions $b_i$ subject to the Wronskian constraint. Expressions for $(\p-\bp)X_i$ can easily be derived. These expressions depend on a constant matrix $N_{ij}$ of size $3$ and determinant $1$; it is not restrictive to assume that $N$ is diagonal with eigenvalues $\nu_1,\nu_2,\nu_3$. 

The rest of the reasoning is similar to the case $W-\bar{W}=0$, with a different formula for the object $\Lambda^{ij}_{mq}$ which appears in eq. (\ref{kijxy}):
\begin{multline}
\Lambda^{ij}_{mq} =(\nu_i-\nu_m)(\nu_j-\nu_q)\left[\left(\frac{2}{X_1^2}-\frac{2\nu_\ell\nu_p}{X_2^2} +\frac{\nu_p-\nu_\ell}{X_1X_2}\right) w_j b_q(w'_ib_m-w_ib'_m) \right.
\\ 
-\left(\frac{2}{X_1^2}-\frac{2\nu_\ell\nu_p}{X_2^2} -\frac{\nu_p-\nu_\ell}{X_1X_2}\right)w_ib_m(w'_jb_q-w_jb'_q) 
\\ \left.
+\frac{\nu_\ell-\nu_p}{X_1X_2}\left(\frac{(\p-\bp)X_1}{X_1}-\frac{(\p-\bp)X_2}{X_2}\right) w_ib_mw_jb_q\right]\ ,
\end{multline}
where the indices $p$ and $\ell$ are such that $\e_{im\ell}$ and $\e_{jqp}$ do not vanish. With such an expression for $\Lambda^{ij}_{mq}$, we find that eq. (\ref{fjkim}) no longer holds, by numerically testing it in various examples of values of $b_i$. This proves that there is no boundary action from which the boundary condition $W+\bar{W}=0$ can be derived.

Remember however that this proof of the non-existence of the boundary action relies on our assumption that only Neumann-type boundary conditions are allowed, and Dirichlet-type boundary conditions do not occur. So there is no contradiction with the special case (\ref{dnc}) when Dirichlet-type conditions could be derived by varying an action (whose boundary term was actually zero). But we saw in Section \ref{secbo} that in the generic case the boundary condition $W+\bar{W}=0$ corresponds to two-dimensional D-branes, and we do not expect Dirichlet conditions to apply.



\begin{thebibliography}{10}
\expandafter\ifx\csname url\endcsname\relax
  \def\url#1{{\tt #1}}\fi
\expandafter\ifx\csname urlprefix\endcsname\relax\def\urlprefix{URL }\fi
\providecommand{\eprint}[2][]{\url{#2}}

\bibitem{fl07c}
V.~A. Fateev, A.~V. Litvinov, {\em Correlation functions in conformal Toda
  field theory I\/}, JHEP 11 p. 002 (2007), \eprint{arXiv:0709.3806 [hep-th]}

\bibitem{zam85}
A.~B. Zamolodchikov, {\em {Infinite Additional Symmetries in Two-Dimensional
  Conformal Quantum Field Theory}\/}, Theor. Math. Phys. 65 pp. 1205--1213
  (1985)

\bibitem{fl88}
V.~A. Fateev, S.~L. Lukyanov, {\em The Models of Two-Dimensional Conformal
  Quantum Field Theory with Z(n) Symmetry\/}, Int. J. Mod. Phys. A3 p. 507
  (1988)

\bibitem{bpz84}
A.~A. Belavin, A.~M. Polyakov, A.~B. Zamolodchikov, {\em Infinite conformal
  symmetry in two-dimensional quantum field theory\/}, Nucl. Phys. B241 pp.
  333--380 (1984)

\bibitem{fz86b}
V.~A. Fateev, A.~B. Zamolodchikov, {\em Conformal quantum field theory models
  in two dimensions having $Z_3$ symmetry\/}, Nucl. Phys. B280 pp. 644--660
  (1987)

\bibitem{fl08}
V.~A. Fateev, A.~V. Litvinov, {\em {Correlation functions in conformal Toda
  field theory II}\/}, JHEP 01 p. 033 (2009), \eprint{0810.3020}

\bibitem{car89}
J.~L. Cardy, {\em Boundary conditions, fusion rules and the Verlinde
  formula\/}, Nucl. Phys. B324 p. 581 (1989)

\bibitem{bs92}
P.~Bouwknegt, K.~Schoutens, {\em W symmetry in conformal field theory\/}, Phys.
  Rept. 223 pp. 183--276 (1993), \eprint{hep-th/9210010}

\bibitem{fl90}
S.~L. Lukyanov, V.~A. Fateev, {\em {Physics reviews: Additional symmetries and
  exactly soluble models in two-dimensional conformal field theory}\/} Chur,
  Switzerland: Harwood (1990) 117 p. (Soviet Scientific Reviews A, Physics:
  15.2)

\bibitem{fms97}
P.~Di~Francesco, P.~Mathieu, D.~Senechal, {\em Conformal field theory\/} New
  York, USA: Springer (1997) 890 p

\bibitem{bw93}
P.~Bowcock, G.~M.~T. Watts, {\em Null vectors, three point and four point
  functions in conformal field theory\/}, Theor. Math. Phys. 98 pp. 350--356
  (1994), \eprint{hep-th/9309146}

\bibitem{rib08b}
S.~Ribault, {\em {On sl3 Knizhnik-Zamolodchikov equations and W3 null-vector
  equations}\/}, JHEP 10 p. 002 (2009), \eprint{0811.4587}

\bibitem{zz95}
A.~B. Zamolodchikov, A.~B. Zamolodchikov, {\em Structure constants and
  conformal bootstrap in Liouville field theory\/}, Nucl. Phys. B477 pp.
  577--605 (1996), \eprint{hep-th/9506136}

\bibitem{ls79}
A.~N. Leznov, M.~V. Saveliev, {\em {Representation of zero curvature for the
  system of nonlinear partial differential equations $x_{\alpha ,z\bar z}={\rm
  exp}(kx)_{\alpha }$ and its integrability.}\/}, Lett. Math. Phys. 3(6) p. 489
  (1979)

\bibitem{af09}
V.~de~Alfaro, A.~T. Filippov, {\em {Multi-exponential models of
  (1+1)-dimensional dilaton gravity and Toda-Liouville integrable models}\/},
  Theor. Math. Phys. 162 pp. 34--56 (2010), \eprint{0902.4445}

\bibitem{fzz00}
V.~Fateev, A.~B. Zamolodchikov, A.~B. Zamolodchikov, {\em Boundary Liouville
  field theory. I: Boundary state and boundary two-point function\/}  (2000),
  \eprint{hep-th/0001012}

\bibitem{tes00}
J.~Teschner, {\em Remarks on Liouville theory with boundary\/}  (2000),
  \eprint{hep-th/0009138}

\bibitem{zz01}
A.~B. Zamolodchikov, A.~B. Zamolodchikov, {\em Liouville field theory on a
  pseudosphere\/}  (2001), \eprint{hep-th/0101152}

\bibitem{fat01}
V.~A. Fateev, {\em Normalization factors, reflection amplitudes and integrable
  systems\/}  (2001), \eprint{hep-th/0103014}

\bibitem{dgg10}
N.~Drukker, D.~Gaiotto, J.~Gomis, {\em {The Virtue of Defects in 4D Gauge
  Theories and 2D CFTs}\/}  (2010), \eprint{1003.1112}

\bibitem{tv78}
L.~Takhtajan, A.~Vinogradov, {\em Theory of the Eisenstein series for the group
  $SL(3,R)$ and its application to the binary problem I\/}, Notes of the LOMI
  seminars 76 p.~5 (1978)

\bibitem{anps94}
A.~Anderson, B.~E.~W. Nilsson, C.~N. Pope, K.~S. Stelle, {\em {The Multivalued
  free field maps of Liouville and Toda gravities}\/}, Nucl. Phys. B430 pp.
  107--152 (1994), \eprint{hep-th/9401007}

\end{thebibliography}

\end{document}